\documentclass[aps, showpacs,preprintnumbers, superscriptaddress, nofootinbibt, twocolumn]{revtex4}
\usepackage{makeidx}
\usepackage{amsfonts}
\usepackage{amsmath}
\usepackage{amssymb}
\usepackage{eurosym}
\usepackage{graphicx}

\def\be{\begin{equation}}
\def\ee{\end{equation}}
\def\bea{\begin{eqnarray}}
\def\eea{\end{eqnarray}}

\usepackage{color}

\begin{document}

\title{Compact stellar structures in Weyl geometric gravity}

\author{Zahra Haghani}
\email{z.haghani@du.ac.ir}
\affiliation{School of Physics, Damghan University, Damghan, 41167-36716, Iran}
\author{Tiberiu Harko}
\email{tiberiu.harko@aira.astro.ro}
\affiliation{Department of Theoretical Physics, National Institute of Physics
and Nuclear Engineering (IFIN-HH), Bucharest, 077125 Romania,}
\affiliation{Department of Physics, Babes-Bolyai University, Kogalniceanu Street,
	Cluj-Napoca 400084, Romania,}
\affiliation{Astronomical Observatory, 19 Ciresilor Street,
	Cluj-Napoca 400487, Romania}


\date{\today }

\begin{abstract}
We consider the structure and physical properties of specific classes of neutron, quark, and Bose-Einstein Condensate stars in the conformally invariant Weyl geometric gravity theory. The basic theory is derived from the simplest conformally invariant action, constructed, in Weyl geometry, from the square of the Weyl scalar, the strength of the Weyl vector, and a matter term, respectively. The action is linearized in the Weyl scalar by introducing an auxiliary scalar field. To keep the theory conformally invariant the trace condition is imposed on the matter energy-momentum tensor. The field equations are derived by varying the action with respect to the metric tensor, Weyl vector field and scalar field.  By adopting a static spherically symmetric interior geometry,  we obtain the field equations, describing the structure and properties of stellar objects in Weyl geometric gravity. The solutions of the field equations are obtained numerically, for different equations of state of the neutron and quark matter. More specifically, constant density stellar models, and models described by the  stiff fluid, radiation fluid, quark bag model, and Bose-Einstein Condensate equations of state are explicitly constructed numerically in both general relativity and Weyl geometric gravity, thus allowing an in depth comparison between the predictions of these two gravitational theories. As a general result it turns out that for all the considered equations of state, Weyl geometric gravity stars are more massive than their general relativistic counterparts. As a possible astrophysical application of the obtained results we suggest that the recently observed neutron stars, with masses in the range of 2$M_{\odot}$ and 3$M_{\odot}$, respectively, could be in fact conformally invariant Weyl geometric neutron or quark stars.
\end{abstract}

\maketitle
\tableofcontents

\section{Introduction}

The structure and the properties of the compact astrophysical objects is a central problem in general relativity. This field of study was initiated already in 1916 by Schwarzschild \cite{Sch1}, who obtained the interior solution for a sphere of constant density, with vanishing pressure on its vacuum boundary. Despite its theoretical simplicity, and limited interest for realistic stellar objects, the Schwarzschild interior solution did attract a lot of interest, and its properties and extensions have been intensively studied \cite{Sch2,Sch3,Sch4,Sch5,Sch6, Sch7}. An important moment in the development of relativistic astrophysics is related to the work by Tolman \cite{T1,T2} and of Oppenheimer and Volkoff \cite{OV}, who obtained the equations of structure of compact general relativistic objects for a static, spherically symmetric geometry. In particular, the equation describing the hydrostatic equilibrium of compact stars was also obtained in these early studies, and it is presently called the Tolman-Oppenheimer-Volkoff (TOV) equation.

The theoretical, as well as the numerical investigations of the TOV equation led to the limiting maximum mass of neutron stars, which was found to be of the order of $3.2M_{\odot}$ \cite{RoRu74}.  This result was obtained by using the principle of causality, the maximally stiff equation of state $p=\rho c^2$, and Le Chatelier's principle,  and it is valid even if the equation of state of matter is unknown in a limited range of densities. On the other hand, Chandrasekhar \cite{Ch} obtained for the limiting mass of white dwarfs the value $M_{Ch}\approx 1.4M_{\odot}$. Theoretical arguments, as well as observational evidence thus led to the assumption, generally accepted for a long time, that neutron stars must have a  mass distribution centered on a value of the order of $1.4M_{\odot}$ \cite{Teu}. This mass value follows from the result that neutron stars must be supported by the neutron degenerate pressure that becomes dominant after the collapse of the white dwarf. The corresponding radius of a $1.4M_{\odot}$ mass neutron star should be of the order of 10 - 15 km, and its average density is of the order of $6\times 10^{14}\;{\rm g/cm^3}$.

However, the standard view on the neutron star masses has changed drastically recently, once more and more exact determinations of the neutron star masses became available \cite{Ho}.  A large number of accurate astronomical observations, as well as the detection of the gravitational waves, has clearly indicated that the masses of neutron stars vary in a much larger range than expected from the simple application of the Chandrasekhar limit. Hence, in \cite{MaMe}, by using combined electromagnetic and gravitational wave information on the binary neutron star merger GW170817, an upper limit of $M_{max}\leq 2.17M_{\odot}$ for the maximum mass of a neutron star was found.  This limit is tighter and less model-dependent than other constraints.  From the analysis of the same event in \cite{Sh} it was pointed out that the neutron matter equation of state has to be sufficiently stiff, implying that the maximum mass of neutron stars has to be much higher than $2M_{\odot}$. This mass value is required so that a long-lived massive neutron star can be formed as the merger remnant for the binary systems of GW170817, for which the initial total mass is greater than $2.73M_{\odot}$. Moreover, since no relativistic optical counterpart was detected, a value of $M_{max} \approx 2.15-2.25M_{\odot}$ can be inferred for the maximum value of the mass of the neutron star. Similar values for the maximum mass of the neutron star have been obtained in \cite{Ru} and \cite{Re}, respectively. Other determinations of the masses of neutron stars, by using the Shapiro delay, gave a mass of the order of $1.928\pm 0.017M_{\odot}$ for the pulsar PSR J1614-2230 \cite{Fo},  and a mass $2.14_{-0.09}^{+0.10}M_{\odot}$ for the Millisecond Pulsar MSP J0740+6620 \cite{Cro}.

Interestingly enough, more intriguing determinations of neutron star masses came from the gravitational event GW190425, which indicated the total mass of the binary neutron stars to be of the order of $3.4M_{\odot}$ \cite{Abb1}, and from GW190408, which did show the possible existence of a neutron star with mass $2.5-2.65M_{\odot}$, merging with a large black hole with mass $26M_{\odot}$ \cite{Abb2}. Hence, these recent observations opened new perspectives of the mass distribution of the neutron stars, making the old paradigm of the standard $1.4M_{\odot}$ mass value untenable.

A possible explanation for the high mass values of the observed compact objects could be obtained by assuming that they contain some exotic components. In \cite{Ko1} it was suggested that rotating strange quark stars in the Color-Flavor-Locked (CFL) phase can have masses in the range $3.8-6M_{\odot}$, and thus they can mimic even stellar mass black holes. The low luminosity of the star makes it difficult to detect. On the other hand, Bose-Einstein Condensate stars, consisting of superfluid condensates with particle masses of the order of two neutron masses, forming Cooper pairs, and scattering length of the order of 10 - 20 fm have maximum masses of the order of $2M_{\odot}$. If the particles composing the condensate are kaons, then the mass of the neutron star can be in the range of $2.4-2.6M_{\odot}$ \cite{Cha}. For a detailed discussion of the astrophysical role of the exotic components, like, for example, quarks and kaon condensates, in the interior of neutron stars see \cite{Gle}.

Important constraints on the nature of the equation of state of the dense matter in the interior of the neutron stars can be obtained by using multi-messenger astronomy. The observations of the binary neutron star merger GW170817, together with the observation of electromagnetic counterparts across the entire spectrum allowed to obtain some important restrictions on the mass ratio $q\leq 1.38$, and the tidal deformability of the source, $\tilde{\Lambda}\geq 197$ \cite{Cou}. These constraints rule out sufficiently soft equations of state of the nuclear matter. Compact binary mergers that include neutron stars lead to the creation of the electromagnetic counterpart, the kilonova, which originate from the neutron-rich outflows from the merger \cite{Ka}. In \cite{Cou} the analysis of the GW170817 event was performed by using standard general relativity.

Another important avenue for the possible explanation of the high mass values of the neutron stars is modified gravity. Modified gravity theories have been proposed mainly to explain the recent accelerating expansion of the Universe (for reviews of the accelerating Universe, dark energy and modified gravity problems see \cite{r1,r2,r3,r4}).

 Modified gravity theories can also open some new windows on the understanding of the structure of compact objects. In modified gravity theories some of the basic stellar structure properties, like the mass-radius relations, maximum masses, or moments or inertia are different as compared with standard general relativity \cite{Olmo}. It is important to point out that even in the case of non-relativistic stars, like white, red or brown dwarfs, modified gravity effects may play an important role in their internal structure, due to the modification of the gravitational interaction inside these astrophysical objects. These modifications are induced through the modified Poisson equation for the gravitational potential, which leads to modifications in the mass, radius, central density, and luminosity of the stars \cite{Olmo}. Moreover,  the Chandrasekhar mass limit for white dwarfs, and the minimum mass for stable hydrogen burning in high-mass brown dwarfs stars are also influenced.  The observational determination of the masses of several neutron stars having values of the order of  $2M_{\odot}$ has led to contradictions  between the predictions of some realistic equations of state of dense matter,  ruling out many of the soft ones (and in particular,  the ones including hyperons) \cite{Olmo}.
Modified  theories of gravity that go beyond general relativity could help in  alleviating or solving  these contradictions, through the corrections that appear in the generalized hydrostatic equilibrium equations of extended stellar models \cite{Olmo}.

There are several extensions of general relativity that could explain the observational properties of the compact astrophysical  objects, like neutron stars, and of their mass distribution, which include $f(R)$,  $f(R,T)$ and, $f\left(R,L_m\right)$ type gravity theories, hybrid-metric Palatini gravity, or theories in which the standard Hilbert-Einstein Lagrangian is extended by a three-form field $A_{\alpha \beta \gamma}$. For a brief review of the astrophysical implications of the above mentioned theories, as well as on the Palatini and metric affine gravity, see \cite{Olmo}, and references therein. In many modified gravity theories, it is possible to obtain stellar mass values for static, spherically symmetric objects, that are very difficult to be obtained in standard General Relativity,  even considering the effects of fast rotation.

One of the important extensions of the Riemann geometry is represented by the Weyl geometry \cite{Weyl, Weyl1}. In Weyl geometry the metric condition is abandoned, and the covariant divergence of the metric tensor in non-zero, $\nabla_{\alpha}g_{\mu \nu}=Q_{\alpha \mu \nu}\neq 0$, where $Q_{\alpha \mu \nu}$ is the non-metricity. In its initial formulation by Weyl, for the non-metricity the particular form $Q_{\alpha \mu \nu}=\omega _\alpha g_{\mu \nu}$ was assumed, where $\omega _\alpha$ is the Weyl vector. For a presentation of the Weyl geometry, and its historical aspects, see \cite{Scholz}. Another important idea introduced by Weyl is the idea of the conformal invariance of physical laws, and the necessity of reformulating Einstein's gravity as a conformally invariant theory. Recently, conformally invariant theories of gravity, and of elementary particle physics, were proposed and discussed in \cite{P1,P1a,P2,P3,P4,G1a,G2}.

A conformally invariant theory of gravity, based on the action $S=-\alpha _g\int{C_{\lambda \mu \nu \kappa}C^{\lambda \mu \nu \kappa}\sqrt{-g}d^4x}=-2\alpha _g\int{\left(R_{\mu \nu}R^{\mu \nu}-R^2/3\right)\sqrt{-g}d^4x}$, where $C_{\lambda \mu \nu \kappa}$ is the (conformally invariant) Weyl tensor, was introduced, and extensively investigated, in  \cite{Ma0,Ma1,Ma2,Ma3,Ma4,Ma5}. In particular, the conformally invariant vacuum field equations do admit an exact solution of the form $ds^2=e^{\nu}dt^2-e^{-\nu}dr^2-r^2d\Omega$, where $e^{\nu}=1-2m/r+\gamma r-\Lambda r^2$, where $m$, $\gamma $, $\Lambda$ are constants \cite{Ma0}.

Weyl geometry, and the corresponding gravitational theory, have found a large number of applications in theoretical physics, astrophysics, and cosmology. Dirac \cite{Dirac1, Dirac2} has proposed an extension of Weyl's theory, which is based on three geometric quantities, the symmetric metric tensor $g_{\mu \nu}$, the Weyl connection vector $\omega _{\mu}$, and the Dirac gauge function $\beta$. The Weyl-Dirac theory was extensively discussed, and applied to cosmology \cite{Rosen,Isrcosm}, and can give a systematic description of the matter creation in the Universe, as well as for its late acceleration. Weyl geometry is the theoretical foundation of the $f(Q)$ type modified gravity theories \cite{Q1, Q2, Q4}, and of its generalizations \cite{Q10, Q17, Q20, Q23, Q24}.  In the $f(Q)$ theory, the basic quantity describing the gravitational field is the (metric dependent) non-metricity $Q$. The action of this theory is given by $S=\int{f(Q)\sqrt{-g}d^4x}$, where $f(Q)$ is an arbitrary function of the non-metricity. An observational investigation of several modified $f(Q)$ models using the redshift approach was performed in \cite{Laz},  by using a variety of observational probes (Type Ia Supernovae, Quasars, Gamma Ray Bursts, Baryon Acoustic Oscillations and Cosmic Microwave Background data). It turns out that $f(Q)$ gravity  provides a credible alternative model for the explanation of the late time acceleration of the Universe. $f(Q)$ type theories have been extended to include the effects of the coupling between non-metricity and matter, described by the matter Lagrangian $L_m$ in \cite{C0a}. This approach was extended to theories with action described by an arbitrary function of the non-metricity and the trace of the energy-momentum tensor $T$, in \cite{C1a,C2a,C3a}, leading to the so-called $f(Q,T)$ theory of gravity. The role of the torsion in the couplings between matter and the basic geometric quantities was investigated in \cite{C4a}. A cosmological model based on $f\left({\cal R}\right)$ gravity, with $f\left({\cal R}\right)=\alpha {\cal R}-\beta {\cal R}^2/2-\gamma /3{\cal R}$ was investigated in \cite{Pinto}, by using the Palatini approach. The model was interpreted in a Weylian geometric framework, and the definitions of the luminosity distance, proper distance, and redshift were extended to Weyl type geometries.

An interesting approach to Weyl gravity, and its applications,  was initiated, and extensively developed, by using a perspective suggested by elementary particle physics, in \cite{Gh1, Gh2,Gh3,Gh4,Gh5,Gh6,Gh7,Gh8}. The basic idea is to linearize in the action of the Weyl quadratic gravity the (geometric) term  $\tilde{R}^2$ with the help of an auxiliary scalar field.  The linearized Weyl quadratic gravity undergoes a spontaneous breaking of $D(1)$ by a geometric, Stueckelberg type, mechanism, with the Weyl gauge field acquiring mass from the spin-zero mode of the $\tilde{R}^ 2$ term in the action. The Stueckelberg mechanism is implemented by setting the scalar field $\phi$ to a constant value, taken as its vacuum expectation value, so that $\phi\rightarrow <\phi>$.  Then, the Weyl vector field  becomes massive, and it absorbs the dynamical scalar field $\phi$, which disappears from the initial scalar-vector-tensor theory. In this way the Einstein-Proca action is recovered from the Weyl action, by eliminating the auxiliary scalar field $\phi$, and thus returning to a vector-tensor theory, as the initial Weyl theory is.

This mode also generates the Planck scale, and the cosmological constant, with the Einstein-Proca action emerging in the broken phase.
Moreover, all mass scales, including the Planck scale, and the cosmological constant, have a geometric origin \cite{Gh8}. Moreover, the Higgs field of the standard model of elementary particle physics has a similar origin, generated by Weyl boson fusion in the early Universe. The coupling of matter and geometry in Weyl gravity was considered in \cite{C1, C2, C3, C4}, with the Palatini formulation of the theory in the presence of geometry-matter coupling was investigated in  \cite{C2}. The quadratic Weyl gravity $\tilde{R}^2+R_{\mu\nu }^2$ in the Palatini formalism was studied in \cite{Gh5}, by assuming that the Weyl connection and the metric are independent. The theory has a spontaneous breaking of gauged scale symmetry, and mass is generated as a purely geometric effect.  The theory leads to a successful inflationary scenario,  by predicting a  tensor-to-scalar ratio $0.007 \leq r \leq 0.01$ for the spectral index $n_s$ (at 95 \% CL) and $N =60$ efolds. A comparative study of inflation in the Weyl quadratic gravity and in the theory obtained in the Palatini approach to the considered action was performed in \cite{Gh6}. These theories have different vectorial non-metricities, induced by the gauge field, and thus lead to different physical predictions. Both theories have a small tensor-to-scalar ratio, $r \sim 10^{-3}$, which is somewhat larger in the Palatini case. The metric Weyl theory gives a dependence $r\left(n_s\right)$ similar to that in the Starobinsky inflation.

It is the goal of the present paper to consider static, spherically symmetric stellar models in the simplest model of Weyl geometric gravity. Our starting point is a gravitational action consisting of the sum of the square $\tilde{R}^2$ of the Weyl scalar, and of the field strength $F_{\mu \nu}^2$ of the Weyl vector. The action can be linearized in the Weyl scalar by introducing an auxiliary scalar field, and then can be reformulated as an effective scalar-vector-tensor theory in Riemann geometry, with the action containing effective couplings between the scalar field and the Ricci scalar, and the Weyl vector field. These terms are conformally invariant by construction.  Moreover, a matter term is also added to the total action. The field equations corresponding to this action are obtained in the metric formalism, by varying the action respect to the metric tensor, Weyl vector and the scalar field.

An important question in conformally invariant gravitational actions is how to implement the conformal invariance of the matter terms. In this work we implement the requirement of the conformal invariance of the matter action by imposing trace condition on the effective matter action $\mathcal{L}_m$, constructed with the help of the ordinary matter action $L_m$, and the square of the Weyl vector $\omega ^2$, so that $\mathcal{L}_m=\mathcal{L}_m\left(L_m,\omega ^2\right)$. Once this condition is satisfied, the corresponding gravitational field equations, and their solutions,  are conformally invariant.

After obtaining the gravitational field equations of the Weyl geometric gravity and the consistency condition in static spherical symmetry, we consider a number of specific stellar models, whose general relativistic counterparts have been intensively investigated. Thus, we investigate constant density, stiff fluid, radiation fluid, quark and Bose-Einstein Condensate stars, by numerically solving the Weyl geometric gravity field equations. In each case a detailed analysis of the astrophysical properties of the stars (mass and density profiles, mass-radius relation, Weyl vector and scalar field behavior) is performed, and the dependence of the stellar structures on the Weyl geometric gravity theory parameters is presented. As a general conclusion of our study it turns out that a larger variety of stellar type objects can be constructed in Weyl
geometric gravity as compared to standard general relativity.

The present paper is organized as follows. In Section~\ref{sect1} we review the basic of Weyl geometry, we introduce the quadratic Weyl geometric gravitational action, and we discuss its linearization in the Weyl scalar. The gravitational field equations are obtained by varying the action with respect to the metric tensor, to the Weyl vector and to the scalar field, together with the consistency condition on the Weyl current. A specific form of the effective matter Lagrangian $\mathcal{L}$ is also adopted, and the corresponding field equations are written down.  In static spherical symmetry the field equations of the Weyl geometric theory are obtained in Section~\ref{sect2}, where a dimensionless representation of the geometric and physical variables is also introduced. Several stellar models are constructed, by numerical solving the gravitational field equations, in Section~\ref{sect3}. Finally, we discuss and conclude our results in Section~\ref{sect4}.

\section{ Weyl geometric gravity in a nutshell}\label{sect1}

 In the present Section we first briefly review the geometrical foundations of the conformally invariant Weyl geometric gravity. The action of the theory is also written down, and linearized in the Weyl, and Ricci curvatures by introducing an auxiliary scalar field. The gravitational field equations are obtained in a general form by varying the action with respect to the metric, and the scalar and Weyl vector field.

\subsection{Geometry, gravitational action, and matter, in conformally invariant Weyl spacetimes}

We begin our presentation of the Weyl geometric gravity theory with a brief review of the Weyl geometry, and of its basic geometrical, and physical quantities. Then the action of Weyl geometric gravity, and the gravitational field equations are presented.

\subsubsection{Recap of Weyl geometry}

 Weyl geometry is constructed as the classes of equivalence $\left\{ g_{\mu
\nu }(x),\omega _{\alpha }(x)\right\} $ of the metric $g_{\mu \nu }(x)$ and of the
Weyl vector gauge field $\omega _{\alpha }(x)$, respectively. These geometric quantities are related by the Weyl gauge
transformations \cite{Gh7},
\be
\tilde{g}_{\mu \nu}=\Omega ^n(x)g_{\mu \nu},\quad \tilde{\omega}_{\alpha}=\omega _{\alpha}-\frac{n}{\alpha}\frac{\partial _\alpha \Omega (x)}{\Omega (x)},
\ee
where $n$ is the Weyl charge.
The Weyl gauge vector field $\omega _{\mu}$ is defined via the Weyl connection $\tilde{\Gamma}
$, which can be obtained from the non-metricity equations,
\begin{equation}\label{5}
\tilde{\nabla}_{\lambda }g_{\mu \nu }=-n\alpha \omega _{\lambda }g_{\mu \nu },
\end{equation}%
where by $\alpha$ we have denoted the Weyl gauge coupling, and
\begin{equation}
\tilde{\nabla}_{\lambda }g_{\mu \nu }=\partial _{\lambda }g_{\mu \nu }-%
\tilde{\Gamma}_{\nu \lambda }^{\rho }g_{\rho \mu }-\tilde{\Gamma}_{\mu
\lambda }^{\rho }g_{\nu \rho }.
\end{equation}

Note that the Weyl geometry is {\it non-metric}. Eq.~(\ref{5}) can be rewritten in an equivalent form as
\begin{equation}
\left( \tilde{\nabla}_{\lambda }+n\alpha \omega _{\lambda }\right) g_{\mu
\nu }=0.
\end{equation}

In the following we will denote by a tilde the geometric and physical quantities defined in Weyl geometry, while the bare quantities represent their Riemannian counterparts. One can construct gauge invariant expressions in a way similar to gauge theory in elementary particle physics, by replacing  the partial derivative by the
Weyl covariant derivative, according to the rule,
\begin{equation}
\partial _{\lambda }\rightarrow \partial _{\lambda }+n
\alpha  \omega _{\lambda }.
\end{equation}

From Eq.~(\ref{5}), by permuting the indices, and combining the resulting relations, we obtain
\begin{equation}
\tilde{\Gamma}_{\mu \nu }^{\lambda }=\Gamma _{\mu \nu }^{\lambda }+\alpha
\frac{n}{2}\left( \delta _{\mu }^{\lambda }\omega _{\nu }+\delta _{\nu
}^{\lambda }\omega _{\mu }-\omega ^{\lambda }g_{\mu \nu }\right) ,
\label{1a}
\end{equation}%
where by
\begin{equation}
\Gamma _{\lambda ,\mu \nu }=\frac{1}{2}\left( \partial _{\nu }g_{\lambda \mu
}+\partial _{\mu }g_{\lambda \nu }-\partial _{\lambda }g_{\mu \nu }\right) ,
\end{equation}%
we have denoted the standard Levi-Civita metric connection of Riemannian geometry, and
\begin{equation}
\tilde{\Gamma}_{\mu \nu }^{\lambda }=g^{\lambda \sigma }\tilde{\Gamma}%
_{\lambda ,\mu \nu }.
\end{equation}

The trace of Eq. (\ref{1a}) gives
\begin{equation}
\tilde{\Gamma}_{\mu }=\Gamma _{\mu }+2n\alpha \omega _{\mu }.
\end{equation}

An important physical and geometrical quantity, the field strength $\tilde{F}_{\mu \nu }$ of the Weyl vector field $\omega _{\mu
}$, is defined as,
\bea
\tilde{F}_{\mu \nu }&=&\tilde{\nabla}_{\mu }\omega _{\nu }-\tilde{\nabla}_{\nu
}\omega _{\mu }=\nabla _{\mu}\omega _{\nu}-\nabla _{\nu}\omega _{\nu}\nonumber\\
&=&\partial _{\mu }\omega _{\nu }-\partial _{\nu }\omega _{\mu
}=F_{\mu \nu}.
\eea

The curvatures in Weyl geometry are calculated, by using the Weyl connection,
according to,
\begin{equation}
\tilde{R}_{\mu \nu \sigma }^{\lambda }=\partial _{\nu }\tilde{\Gamma}_{\mu
\sigma }^{\lambda }-\partial _{\sigma }\tilde{\Gamma}_{\mu \nu }^{\lambda }+%
\tilde{\Gamma}_{\rho \nu }^{\lambda }\tilde{\Gamma}_{\mu \sigma }^{\rho }-%
\tilde{\Gamma}_{\rho \sigma }^{\lambda }\tilde{\Gamma}_{\mu \nu }^{\rho },
\end{equation}%
and
\begin{equation}
\tilde{R}_{\mu \nu }=\tilde{R}_{~\mu \lambda \nu }^{\lambda },\quad
\tilde{R}%
=g^{\mu \sigma }\tilde{R}_{\mu \sigma },
\end{equation}%
respectively. However,  the Weyl curvature does not have the symmetry properties of the Riemannian curvature tensor, but instead it satisfies the condition
\be
\tilde{R}_{\mu \nu \lambda \sigma}=-\tilde{R}_{\nu \mu \lambda \sigma}-\tilde{F}_{\lambda \sigma}g_{\mu \nu},
\ee
giving,
\be\label{F}
\tilde{R}_{\mu \nu}-\tilde{R}_{\nu \mu}=2\tilde{F}_{\nu \mu},
\ee
a relation that gives the geometrical interpretation of the strength of the Weyl vector.

The Weyl scalar can be expressed in terms of Riemannian geometric quantities as
\begin{equation}
\tilde{R}=R-3n\alpha \nabla _{\mu }\omega ^{\mu }-\frac{3}{2}\left( n\alpha\right) ^{2}\omega _{\mu }\omega ^{\mu }.  \label{R}
\end{equation}

The Weyl scalar $\tilde{R}$ transforms covariantly, while $\sqrt{-g}\tilde{R}^{2}$ is
invariant with respect to the conformal transformations.
%
%

In the following we will restrict our investigations to the case in which the  Weyl charge $n$ takes the value $n=2$ only.  Moreover, we would like to point out that in the present approach to Weyl gravity we adopt a purely geometric, that is, we do not consider the basic quantities in Weyl geometry as having a direct physical meaning. This also refers to the field strength F, which we interpret as a purely geometric quantity, defined by Eq.~(\ref{F}). The physical role of $F_{\mu \nu}$  is indirect, as a geometric contribution to the total energy-momentum balance of a gravitational system. Hence, the possible physical role or interpretation  of $F_{\mu \nu}$ must be considered on a case by case basis, in the given cosmological/astrophysical context. On the other hand, we assume that $F_{\mu \nu}$  is "physical" in the sense that it is geometry acting on matter.

\subsubsection{Action of the Weyl geometric gravity}

Subsequently,  we will consider the simplest Weyl geometric type, conformally invariant gravity theory. To describe the properties of the gravitational field, by using only the two basic scalars of the Weyl geometry $\left( \tilde{R},\tilde{F}%
_{\mu \nu }^{2}\right) $, the following
action was initially proposed by Weyl \cite{Weyl, Weyl1}, and recently reconsidered in \cite{Gh7},
\begin{eqnarray}\label{S1}
S=\int \Big[\frac{1}{4!\xi ^2}\tilde{R}^{2}-\frac{1}{4}\,\tilde{F}_{\mu
	\nu }^{2}\Big]\sqrt{-g}d^{4}x+S_m,
\end{eqnarray}
where $\xi$ is a coupling constant. In the action (\ref{S1}) we have also introduced  the effective matter action $S_m$,
\be\label{M1}
S_m=\beta \int \mathcal{L}_{m}\sqrt{-g}d^4x,
\ee
where $\beta$ is a constant, and the effective matter Lagrangian $\mathcal{L}_m$,
\be
\mathcal{L}_m=\mathcal{L}_m(L_m,\omega^2,\psi)
\ee
in general can depend on the ordinary matter Lagrangian $L_m$, on the Weyl vector through $\omega^2=\omega^\mu \omega_\mu$, and on the matter fields $\psi$. In fact, $\mathcal{L}_m$ may also contain different couplings between ordinary baryonic matter, the Weyl vector, and the fields $\psi$.

We introduce at this moment an auxiliary scalar field $\phi$, according to the definition \cite{Gh7},
\begin{equation}
\tilde{R}^{2}\equiv 2\phi^{2}\tilde{R}-\phi^{4}.  \label{R2}
\end{equation}
We then substitute $\tilde{R}^{2}\rightarrow2\phi^{2}\tilde{R}-\phi ^{4}$ into the geometric part of the action in  Eq.~\eqref{S1}.
The variation of the action (\ref{S1}) with respect to $\phi _{0}$ leads to
the equation
\begin{equation}
\phi\left( \tilde{R}-\phi ^{2}\right) =0,
\end{equation}%
which fixes $\phi^{2}$ as
\begin{equation}
\phi^{2}=\tilde{R}.
\end{equation}%
Thus, through this identification, we recover the original form of the
Lagrangian, as defined in the initial Weyl geometric action. Hence, the $\tilde{R}^2$ type Weyl geometric gravitational models have the remarkable property of allowing their linearization in the Ricci scalar via the introduction of the scalar degree of freedom.

It should be mentioned that the geometrical part of the action (\ref{S1}) is invariant under the transformations,
\begin{align}
\hat{g}_{\mu\nu}=\Omega^2 g_{\mu\nu},\quad \hat{\omega}_\mu=\omega_\mu-\frac{1}{\alpha}\partial_{\mu}\ln(\Omega^2), \quad \hat{\phi}=\frac{\phi}{\Omega}.
\end{align}

The matter part of the  action (\ref{S1}), Eq.~(\ref{M1}), should not be necessarily  gauge invariant, but its variation must be \cite{Bera,Berb,Berc,Berd}. By taking the variation of the matter action (\ref{M1}), we obtain
\bea
\delta S_{m}&=&-\frac{1}{2}\int T^{(tot)}_{\mu \nu}\delta g^{\mu \nu }\sqrt{-g}%
d^{4}x+\int G^{\mu }\delta \omega _{\mu }\sqrt{-g}d^{4}x\nonumber\\
&&+\int \frac{\delta
L_{m}}{\delta \psi }\delta \psi ,
\eea
where $T^{(tot)}_{\mu\nu}$ is the effective total energy-momentum tensor, defined as
\be
T^{(tot)}_{\mu\nu}=-\frac{2}{\sqrt{-g}}\frac{\delta\left( \sqrt{-g} \mathcal{L}_m\left(L_m,\omega ^2,\psi\right)\right)}{\delta g^{\mu\nu}},
\ee
and
\be
G^{\mu}=\frac{\delta \mathcal{L}_m\left(L_m,\omega ^2,\psi\right)}{\delta \omega _{\mu}},
\ee
is the Weyl current \cite{Bera,Berb,Berc,Berd}. In the following we assume that $\delta \mathcal{L}_m/\delta \psi =0$.

By taking into account that \cite{Bera,Berb,Berc,Berd}
\begin{equation}
\delta g^{\mu \nu }=\frac{2\delta \Omega }{\Omega ^{3}}\tilde{g}^{\mu \nu
}=2\frac{\delta \Omega }{\Omega }g^{\mu \nu },
\end{equation}
\bea
\delta \omega _{\mu }&=&\frac{2}{\alpha }\delta \frac{\partial _{\mu }\Omega
}{\Omega }=\frac{2}{\alpha }\delta \left( \partial _{\mu }\ln \Omega
\right) =\frac{2}{\alpha }\partial _{\mu }\left( \delta \ln \Omega \right)\nonumber\\
&=&\frac{2}{\alpha }\partial _{\mu }\left( \frac{\delta \Omega }{\Omega }%
\right)=\frac{2}{\alpha }\nabla _{\mu }\left( \frac{\delta \Omega }{\Omega }%
\right) ,
\eea
for the variation of the matter action we obtain the condition,
\bea\label{Cond}
\delta S_{m}&=&-\int T^{(tot)}_{\mu \nu} g^{\mu \nu }\frac{\delta \Omega }{\Omega }%
\sqrt{-g}d^{4}x \nonumber\\
&&+\frac{2}{\alpha }\int G^{\mu }\nabla _{\mu }\left( \frac{%
\delta \Omega }{\Omega }\right) \sqrt{-g}d^{4}x=0.
\eea

After a partial integration in the above equation, and by using the Gauss theorem, from Eq.~(\ref{Cond}) we obtain the consistency (trace) condition
\begin{align}\label{tr}
T^{(tot)}=-\frac{4}{\alpha}\nabla_\mu\left(\omega^\mu \frac{\partial\mathcal{L}_m\left(L_m, \omega ^2, \psi\right)}{\partial \omega^2}\right),
\end{align}
where $T^{(tot)}=g^{\mu\nu}T^{(tot)}_{\mu\nu}$ is the trace of the effective energy-momentum tensor, constructed with the help of the effective matter Lagrangian $\mathcal{L}_m$.

Eq.~\eqref{tr} shows that when $\mathcal{L}_m=L_m$, this constraint leads to the familiar form $T^{(m)}=0$, where $T^{(m)}$ is the trace of the ordinary matter energy-momentum tensor,  which means that the only conformally invariant matter has a traceless energy-momentum tensor.

By substituting Eqs.~\eqref{R} and (\ref{R2}) into the action (\ref{S1}),  we obtain
\begin{align}\label{sf}
S=\beta \int d^4 x \sqrt{-g} \bigg[&\frac{1}{2\kappa^2}\left( R -6 \alpha^2 \omega_\mu \omega^\mu -12\xi^2\bar{\phi}^2\right)\bar{\phi}^2\nonumber\\&-\frac14 F_{\mu\nu}^2+\mathcal{L}_m \bigg],
\end{align}
where $\bar{\phi}=\phi/\xi$ and $\kappa^2=6\beta$. Note that with these redefinitions, the scalar field $\bar{\phi}$ is dimensionless in the unites $c=1$ and $\kappa^2=1$. To have a canonical kinetic term for the Weyl vector, we have used the redefinitions $\omega_\mu\rightarrow\sqrt{\beta}\omega_\mu$ and $\alpha\rightarrow\alpha/\sqrt{\beta}$.
 It also should be mentioned that we have imposed the gauge condition  $\nabla_\mu\omega^\mu=0$. With the use of the gauge condition, the trace condition (\ref{tr}) becomes
 \begin{align}
T^{(tot)}=-\frac{4}{\alpha}\omega^\mu \nabla_\mu \left[\frac{\partial\mathcal{L}_m\left(L_m, \omega ^2, \psi\right)}{\partial \omega^2}\right].
\end{align}

When $\mathcal{L}_m=L_m$, with $L_m$ independent of $\omega ^2$, and on the other matter fields $\psi$,  it follows that $\partial L_m/\partial \omega ^2\equiv 0$, and therefore in this case we also recover the condition $T^{(tot)}=T^{(m)}=0$, that is, in Weyl geometric gravity only radiative type matter sources give conformally invariant field equations without imposing any supplementary conditions.

\subsection{Gravitational field equations}

In the following, we will assume for the effective matter Lagrangian $\mathcal{L}_m$ the simple form,
\be\label{Lm1}
\mathcal{L}_m=L_m+\gamma \omega^2,
\ee
where $\gamma$ is a constant. The supplementary term added to the matter action is necessary to assure the conformal invariance of the theory, and to satisfy the trace condition.

By varying the gravitational action (\ref{sf}) with respect to the Weyl vector $%
\omega_{\mu}$, it follows that $\omega _{\mu}$ satisfies {\it the generalized system of
Maxwell-Proca type equations},
\begin{equation}
\nabla_\nu F^{\mu\nu}+\frac{3\alpha^2}{2\kappa^2}\,\bar{\phi}\,\omega^\mu-2\gamma\, \omega^\mu=0,
\end{equation}

Due to its antisymmetry, the Weyl field strength $\tilde{F}^{\mu \nu }$
satisfies automatically, in Riemann geometry,  the equations
\begin{equation}\label{Proca2}
\nabla _{\sigma }F_{\mu \nu }+\nabla _{\mu }F_{\nu \sigma }+\nabla
_{\nu }F_{\sigma \mu }=0.
\end{equation}

By varying the action (\ref{sf}) with respect to the metric tensor, and by introducing the Einstein tensor, the gravitational field equations of the Weyl geometric gravity can be written as
\begin{align}\label{feq1}
\bar{\phi}^2 & G_{\mu\nu}-\kappa^2\left(T_{\mu\nu}+F_\mu ^{~\alpha} F_{\nu\alpha}-\frac14 F_{\mu\nu}F^{\mu\nu}\right)+6\xi^2\bar{\phi}^4 g_{\mu\nu}\nonumber\\&+\left(\gamma\kappa^2-\frac{3}{4}\alpha^2\bar{\phi}^2\right)\left(2\omega_\mu \omega_\nu-g_{\mu\nu} \omega^2\right)+g_{\mu\nu}\Box
\bar{\phi}^2\nonumber\\&-\nabla
_\nu \nabla_\mu \bar{\phi}^2=0.
\end{align}

The field equation of the scalar field becomes
\begin{align}\label{phi}
\bar{\phi}^2=\frac{1}{48\xi^2}\left(2R-3\alpha^2 \omega^2\right).
\end{align}

For the choice (\ref{Lm1}) of the effective matter Lagrangian,  the constraint equation \eqref{tr} takes the form,
\begin{align}\label{tr1}
T^{(m)}=-2\gamma \omega^2.
\end{align}

\section{Static spherically symmetric field equations}\label{sect2}

In the present Section we present the static, spherically symmetric gravitational field equations, describing the interior structure of compact Weyl geometric type stars.
We assume that the interior line element in coordinates $\left(t,r,\theta,\phi\right)$ is given by the standard expression,
\begin{align}
ds^2=-e^{-2f(r)}dt^2+\frac{1}{g(r)}dr^2 +r^2 d\Omega^2,
\end{align}
where $f(r)$ and $g(r)$ are arbitrary functions of the radial coordinate $r$, and $d\Omega ^2=d\theta ^2+\sin ^2\theta d\phi^2$. Moreover, we represent $g(r)$ as,
\begin{align}
g(r)=1-\frac{2 \,m(r)}{r},
\end{align}
where $m(r)$ has the physical interpretation as the total (effective) mass of the star.
For the Weyl vector we assume the form,
\begin{align}
\omega^\mu=e^{f(r)}h(r)\delta^\mu_t,
\end{align}
where $h(r)$ is a function to be determined from the field equations. With be above assumptions for the metric and the vector field $\omega_{\mu }$, the  gauge condition, $\nabla _{\mu}\omega ^{\mu}=0$, is automatically satisfied.

For the Lagrangian of the ordinary matter we adopt the expression
\begin{align}
L_m=-\rho,
\end{align}
where $\rho$ is the energy density of the baryonic matter.  The energy momentum tensor of the ordinary matter is given by
\begin{align}
T^m_{\mu\nu}=\left(p+\rho \right)u_\mu u_\nu +p\, g_{\mu\nu},
\end{align}
where by $p$ we have denoted the thermodynamic pressure of the matter, while $u_{\mu}$ is the matter four-velocity.

The non-zero component of the Weyl vector field equation is obtained as
\begin{align}
h''+&h' \left(\frac{2}{r}+\frac{g'}{2
	g}-f'\right)+h \left(-f''-\frac{2 f'}{r}+\frac{2 \gamma}{g} -\frac{f' g'}{2g}\right)\nonumber\\&-\frac{3 \alpha ^2 }{2 \kappa ^2 g} \bar{\phi }^2 h=0.
\end{align}

The $(00)$ component of the metric field equation (\ref{feq1}) is given by,
\begin{align}
\frac{1}{r^2}&\left(1-r g'-g\right)\bar{\phi}^2=3\bar{\phi}^2( 2 \xi ^2 \bar{\phi }^2+\frac{1}{4} \alpha ^2 h^2)\nonumber\\&+2g \left[\bar{\phi }
\left(\frac{2 \bar{\phi }'}{r}+ \bar{\phi }''\right)+ \bar{\phi
}'^2\right]+ g' \bar{\phi }\bar{\phi }'+\kappa^2 \rho\nonumber\\&+\kappa ^2 \left[\frac{1}{2}h^2 \left( g
f'^2-2\gamma \right)-\frac{1}{2} gh'  (2 h\,f' - h')
\right].
\end{align}

The spatial components of the metric field equation (\ref{feq1}) are obtained as,
\begin{align}
\frac{1}{r^2}&\left(1-g+2 r g f'\right)\bar{\phi}^2=-\frac{2}{r} g \left(r f'-2\right)\bar{\phi } \bar{\phi }'-\kappa^2 p \nonumber\\&+ 3( 2 \xi ^2 \bar{\phi }^2-\frac{1}{4} \alpha ^2 h^2)\bar{\phi}^2+\kappa ^2 \left[\frac{1}{2} g \left(h'-h
f'\right)^2+\gamma  h^2\right],
\end{align}
and
\begin{align}
 g \bar{\phi}^2&f''+\frac{1}{2 r}\left(r f'-1\right) \left(g'-2 g f'\right)\bar{\phi}^2=3\bar{\phi }^2( 2 \xi ^2 \bar{\phi }^2-\frac{1}{4} \alpha ^2 h^2 )\nonumber\\&
 +2g \left[\bar{\phi } \bar{\phi }' \left(\frac{1}{r}- f'\right)+ \bar{\phi}'^2+ \bar{\phi } \bar{\phi }''\right]+g'\bar{\phi } \bar{\phi }'-\kappa^2 p\nonumber\\&
 +\kappa ^2 \left[\gamma  h^2-\frac{1}{2} g
 \left(h'-h f'\right)^2\right],
\end{align}
respectively. The non-conservation (balance) equation of the energy momentum tensor is given by
\begin{align}
&\kappa ^2 p' +\frac{3\alpha ^2}{2}  h \bar{\phi }^2 \left(h'-h f'\right)+\bar{\phi } \bar{\phi }'  \left[g'\left(f'-\frac{2}{r}\right)-24 \xi ^2 \bar{\phi }^2\right]\nonumber\\&+\bar{\phi } \bar{\phi }' \left[g
\left(2 f''-\frac{2 \left(r f'-1\right)^2}{r^2}\right)+\frac{3 \alpha ^2
	h^2}{2}+\frac{2}{r^2}\right]\nonumber\\&-\frac{1}{2} \kappa ^2 \left[2 f' \left(-2 \gamma  h^2+p+\rho
\right)+g' \left(h'-h f'\right)^2+4 \gamma  h h'\right]\nonumber\\&-\frac{\kappa ^2}{r} g \left(h
	f'-h'\right) \left[h r f''+f' \left(r h'+2 h\right)-r h''-2 h'\right]=0.
\end{align}

The scalar field equation \eqref{phi}, and the trace constraint equation \eqref{tr1} can be written as,
\begin{align}
24 \xi ^2& \bar{\phi }^2+ g' \left(\frac{2}{r}- f'\right)-\frac{2}{r^2}+\frac{ 2 g}{r^2} \left(r
	f'-1\right)^2\nonumber\\&-2 g f''-\frac{3}{2} \alpha ^2 h^2=0,
\end{align}
and
\begin{align}\label{tr3}
3 p-\rho=2 \gamma  h^2,
\end{align}
respectively. The trace condition allows us to obtain immediately the initial conditions at the center of the star for the Weyl vector $h$, which are given by
\be
h^2(0)=\frac{1}{2\gamma}\left(3p(0)-\rho (0)\right).
\ee

In the next Sections, in order to obtain the numerical solutions of the field equations, we use a set of dimensionless parameters and variables for the geometrical and physical quantities,  which are defined as
\begin{align}
\bar{p}=\frac{p}{\rho_c},\quad \bar{\rho}=\frac{\rho}{\rho_c}, \quad \bar{m}=\sqrt{\rho_c}m,\quad \eta=\sqrt{\rho_c}\, r, \nonumber\\
\bar{\gamma}=\frac{\gamma}{\rho_c},\quad \bar{\alpha}=\frac{\alpha}{\sqrt{\rho_c}}, \quad  \bar{\xi}=\frac{\xi}{\sqrt{\rho_c}}.\qquad
\end{align}

In the calculations we set $\rho_c=2.45\times10^{15}\, {\rm g/cm}^3$.

\section{Stellar models in Weyl geometric gravity}\label{sect3}

In the present Section we will consider several specific stellar models in conformally invariant Weyl geometric gravity. In particular, we will consider constant density stars, stiff fluid stars, described by the Zeldovich equation of state, photon stars, quark stars, and Bose-Einstein Condensate stars, respectively. In all cases the static spherically symmetric field equations of the Weyl geometric gravity are integrated numerically, with the consistency condition of the Weyl current fully satisfied.

\subsection{Constant density stars}

The first interior solution for a static fluid sphere in general relativity was obtained by Schwarzschild \cite{Sch1}, who considered
the simple case of a static, spherically symmetric fluid sphere having a uniform density $\rho$. Even that models of constant density stars are considered unphysical, since, for example, they have a finite density, and a boundary, and an infinite speed of sound, they are very useful theoretical tools as a limiting case for the understanding of the structure of compact objects. Moreover, they can also be used for an approximate description of neutron stars.
 We consider now constant density stars in Weyl geometric gravity, with the matter density satisfying the condition
\begin{align}
\rho=\rho_0={\rm constant}.
\end{align}
In this case, by using the condition of the constant density of the star in the consistency condition Eq.~\eqref{tr3}, we obtain the constraint,
\begin{align}\label{hcd}
h^2=\frac{1}{2\gamma}\left(3p-\rho_0\right).
\end{align}

\begin{figure*}[htbp]
	\centering
	\includegraphics[width=8.0cm]{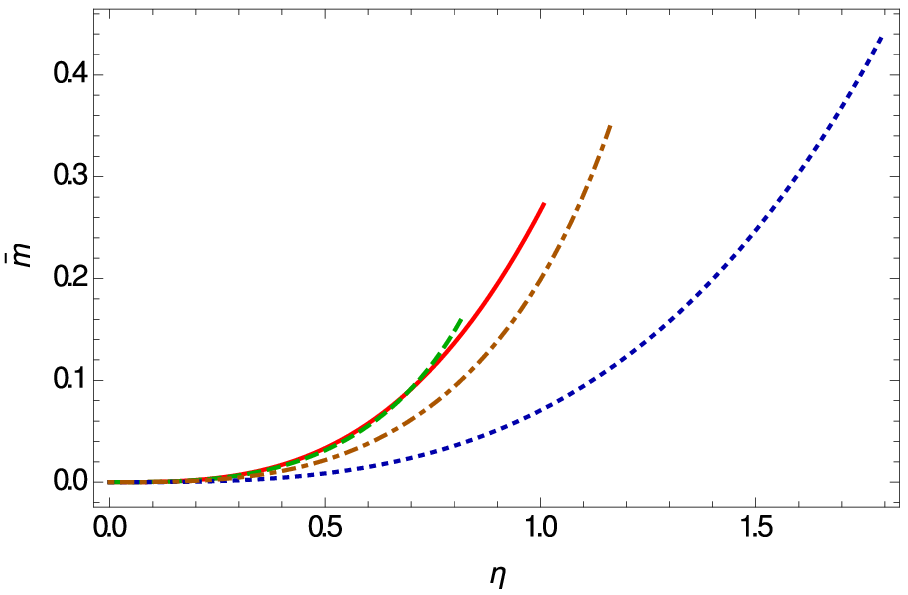}\hspace{.4cm}
	\includegraphics[width=8.0cm]{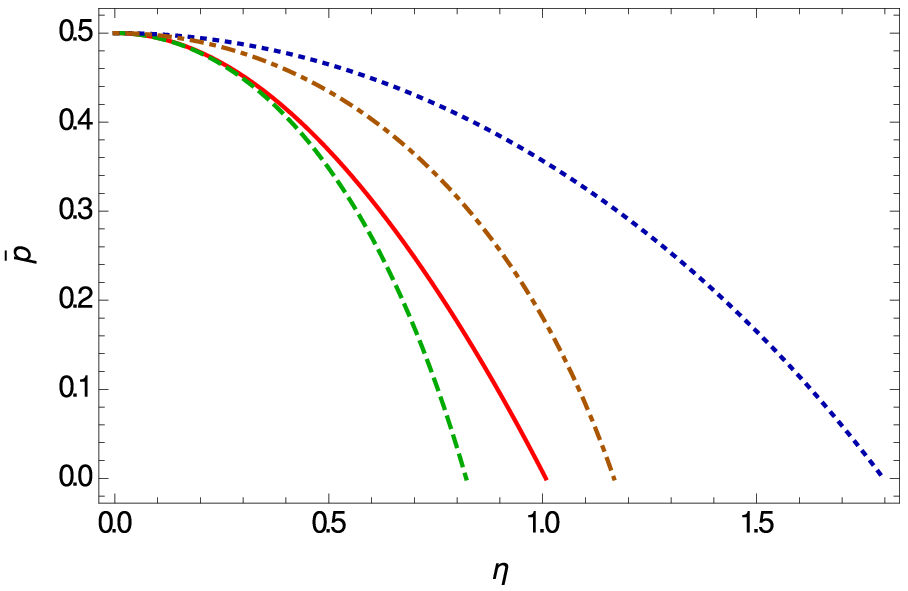}
	\caption{Variation of the interior mass (left panel), and pressure (right panel) profiles of constant density  stars in Weyl geometric gravity as a function of the radial distance from the center of the  star $\eta$ for three different values of the constants $\bar{\alpha}$, $\bar{\xi}$ and $\bar{\gamma}$: $\bar{\alpha}=0.4$, $\bar{\xi}=0.19$  and  $\bar{\gamma}=-0.2$  (dashed curve), $\bar{\alpha}=0.0$,  $\bar{\xi}=0.07$  and  $\bar{\gamma}=-0.2$  (dotted curve), and $\bar{\alpha}=0.2$,  $\bar{\xi}=0.59$  and  $\bar{\gamma}=-0.1$ (dot-dashed curve). The solid curve represents the standard general relativistic  mass and density profile for the constant density stars. The central pressure for all cases is $1.2\times 10^{15} \;{\rm g/cm}^3$.}
	\label{consdens-mass-dens}
\end{figure*}

Eq.~(\ref{hcd}) determines the initial conditions at the center of the star as $h^2(0)=(1/2\gamma) \left(3p(0)-\rho_0\right)$. The mass and pressure profiles of the constant density stars in Weyl geometric gravity are represented, for different values of the coupling constants,  in Fig.~\ref{consdens-mass-dens}. For the numerical solution we have considered pressures in the range $4.6\times 10^{12}\; {\rm g/cm}^3$ and $1.2\times 10^{15}\; {\rm g/cm}^3$, respectively. The density inside the star is $\rho_0=3.9\times 10^{15}\;{\rm g/cm}^3$.

The pressure is a monotonically decreasing function of the radial coordinate, and it vanishes for a given $r=R$, which allows to uniquely define a stellar radius, and a stellar surface. The radius of the star depends significantly on the coupling constants of the model, both larger and smaller radii being allowed. The constant density star becomes more massive in Weyl geometric gravity, as compared to the general relativistic case. Since the pressure is a decreasing function inside the star, and finally becomes zero at the surface star,  from Eq.~\eqref{hcd} it follows that in order to have positive values of $h^2$ inside the star, we should have $3p_c < \rho_0$, where $p_c$ is the central pressure, and $\bar{\gamma}<0$, respectively.

\begin{figure*}[htbp]
	\centering
	\includegraphics[width=8.0cm]{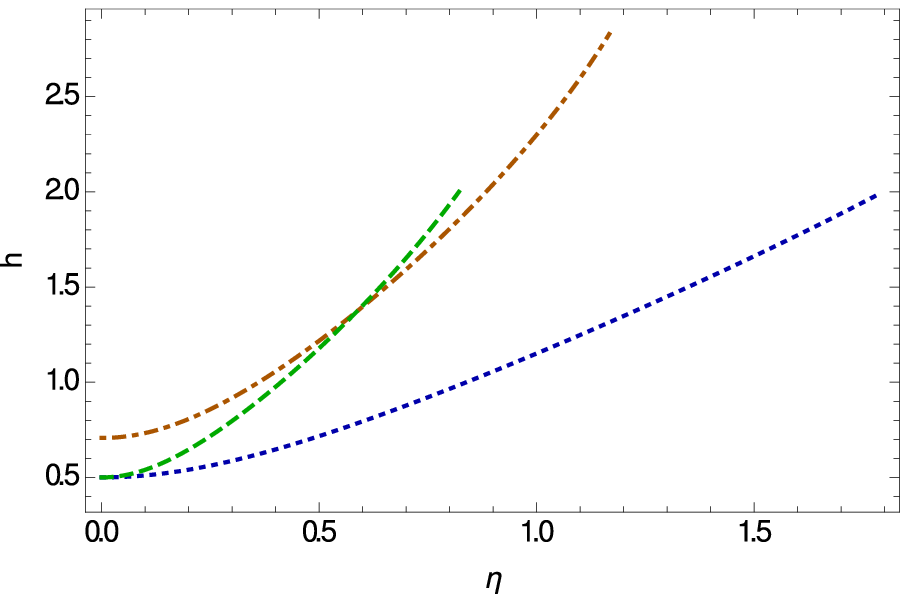}\hspace{.4cm}
	\includegraphics[width=8.0cm]{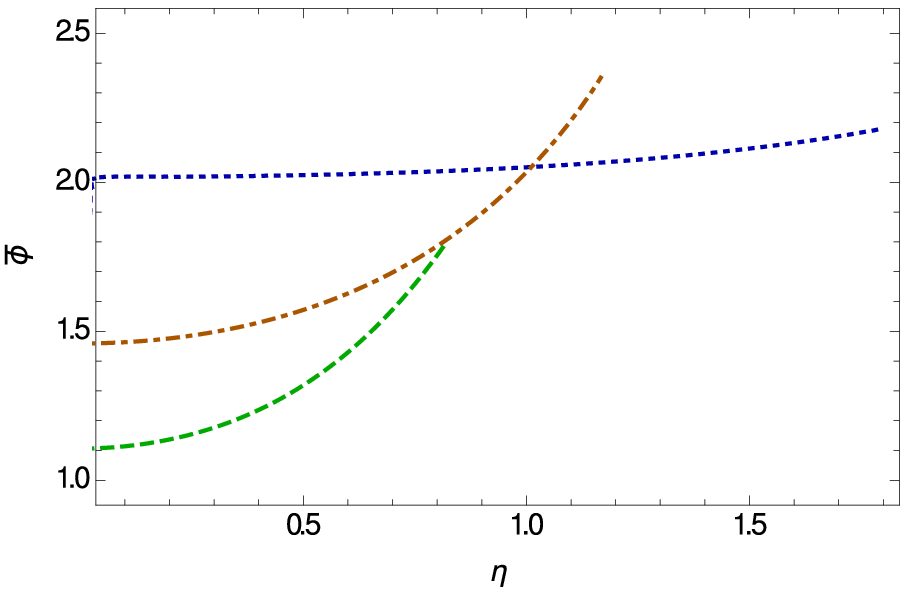}
	\caption{Variation of the scaled temporal component of Weyl vector field $h$ (left panel) and the scalar field $\bar{\phi }$ (right panel) inside the constant density  star in Weyl geometric gravity as a function of the radial distance from the center of the  star $\eta$ for three different values of the
		constants$\bar{\alpha}$, $\bar{\xi}$ and $\bar{\gamma}$: $\bar{\alpha}=0.4$, $\bar{\xi}=0.19$  and  $\bar{\gamma}=-0.2$  (dashed curve), $\bar{\alpha}=0.0$,  $\bar{\xi}=0.07$  and  $\bar{\gamma}=-0.2$  (dotted curve), and $\bar{\alpha}=0.2$,  $\bar{\xi}=0.59$  and  $\bar{\gamma}=-0.1$ (dot-dashed curve). The central pressure for all cases is $1.2\times 10^{15}\; {\rm g/cm}^3$.}
	\label{consdens-vec}
\end{figure*}

The variations of the temporal component of the Weyl vector, and of the scalar field, are represented in Fig.~\ref{consdens-vec}. Both $h$ and $\bar{\phi}$ are increasing functions of $\eta$, and they reach their maximum values on the surface of the star.

\begin{figure}[htbp]
	\centering
	\includegraphics[width=8.0cm]{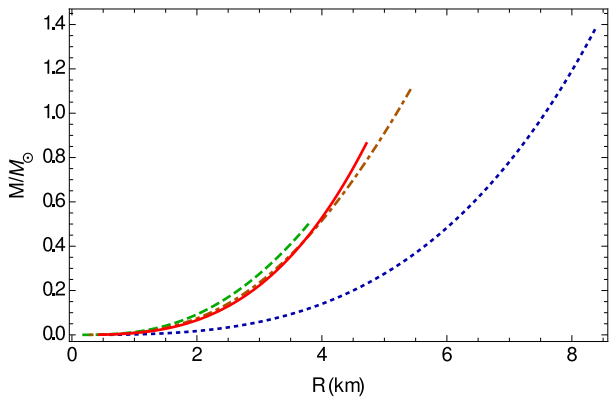}
	\caption{The mass-radius relation for constant density stars in Weyl geometric gravity for three different values of the	constants $\bar{\alpha}$, $\bar{\xi}$ and $\bar{\gamma}$: $\bar{\alpha}=0.4$, $\bar{\xi}=0.19$  and  $\bar{\gamma}=-0.2$  (dashed curve), $\bar{\alpha}=0.0$,  $\bar{\xi}=0.07$  and  $\bar{\gamma}=-0.2$  (dotted curve), and $\bar{\alpha}=0.2$,  $\bar{\xi}=0.59$  and  $\bar{\gamma}=-0.1$ (dot-dashed curve). The solid curve represents the standard general relativistic  mass-radius relation for constant density stars. }
	\label{consdens-mr}
\end{figure}

The mass - radius relation for constant density stars in Weyl geometric gravity is represented in Fig.~\ref{consdens-mr}, for different values of the model parameters. Similarly to the Newtonian case, there is no limiting maximum mass for constant density configurations. However, the existence of more massive stellar structures in Weyl geometric gravity are possible, with significant increases in the mass and radius of the constant density star.

\subsection{Stiff fluid stars}

The stiff fluid (Zeldovich) causal equation of state plays an important role in stellar astrophysics. It was used in \cite{RoRu74}, together with Le Chatelier's principle, to obtain the basic astrophysical result that the maximum mass of a stable neutron star cannot exceed $3.2M_{\odot}$. This extremal value of the mass  is valid even if the equation of state of the dense matter is unknown in a limited range of densities. The existence of an absolute maximum mass of a neutron star provides a powerful method for observationally distinguishing neutron stars from black holes.

We investigate now stiff fluid stars in Weyl geometric gravity, with the equation of state given by,
\begin{align}
	p=\rho.
\end{align}

We can obtain the temporal component of the Weyl vector inside the star using Eq.~\eqref{tr3}, and the stiff fluid equation of state, which together give,
\begin{align}
	h^2=\frac{1}{\gamma}\,\rho.
\end{align}

Hence, in order to satisfy the Weyl consistency condition we must choose positive values for the constant model parameter $\gamma$. The initial condition for the Weyl vector component is $h^2(0)=\rho_c/\gamma$.  In our numerical investigations
the stop point in the integration is $\rho=2.7\times 10^{14}\;{\rm g/cm}^3$. The range of the central densities in the numerical solution is between $3.2\times 10^{14}\, {\rm g/cm}^3$, and $1.1\times 10^{16}\, {\rm g/cm}^3$. In this case, the maximum mass of the standard general relativistic stars is $M=2M_\odot$, with radius $R=11.28\,{\rm km}$, and the central density $\rho_c=3.27\times 10^{15}\,{\rm g/cm}^3$.

\begin{figure*}[htbp]
	\centering
	\includegraphics[width=8.0cm]{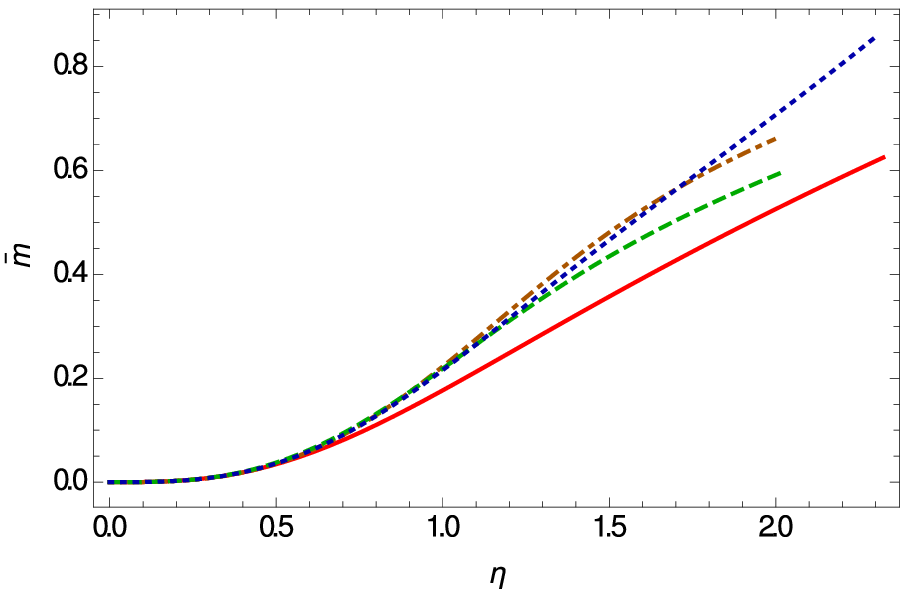}\hspace{.4cm}
	\includegraphics[width=8.0cm]{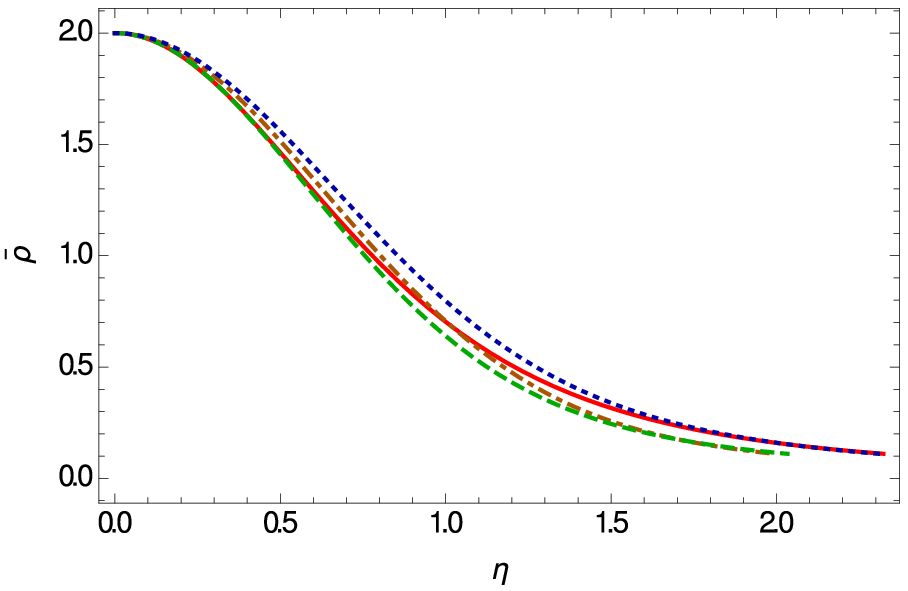}
	\caption{Variation of the interior mass (left panel) and density (right panel) profiles of stiff fluid stars in Weyl geometric gravity as a function of the radial distance from the center of the  star $\eta$ for three different values of the constants $\bar{\alpha}$, $\bar{\xi}$ and $\bar{\gamma}$: $\bar{\alpha}=0.5$, $\bar{\xi}=0.05$  and  $\bar{\gamma}=0.76$  (dashed curve), $\bar{\alpha}=0.55$,  $\bar{\xi}=0.59$  and  $\bar{\gamma}=1.1$  (dotted curve), and $\bar{\alpha}=0.3$,  $\bar{\xi}=0.19$  and  $\bar{\gamma}=0.31$ (dot-dashed curve).  The solid curve represents the standard general relativistic  mass and density profile for stiff stars. For all cases the central density is $4.9 \times 10^{15}\; {\rm g/cm}^3$. }
	\label{stiff-mass-dens}
\end{figure*}

The variations of the mass and density profiles of the stiff fluid stars in Weyl geometric gravity are represented, for different values of the model parameters, in Fig.~\ref{stiff-mass-dens}. The density vanishes on the vacuum boundary of the star, which allows to define a unique radius of the star. The mass profile indicates the possibility of the important increase in the mass of then star, due to the presence of the Weyl geometric effects.

\begin{figure*}[htbp]
	\centering
	\includegraphics[width=8.0cm]{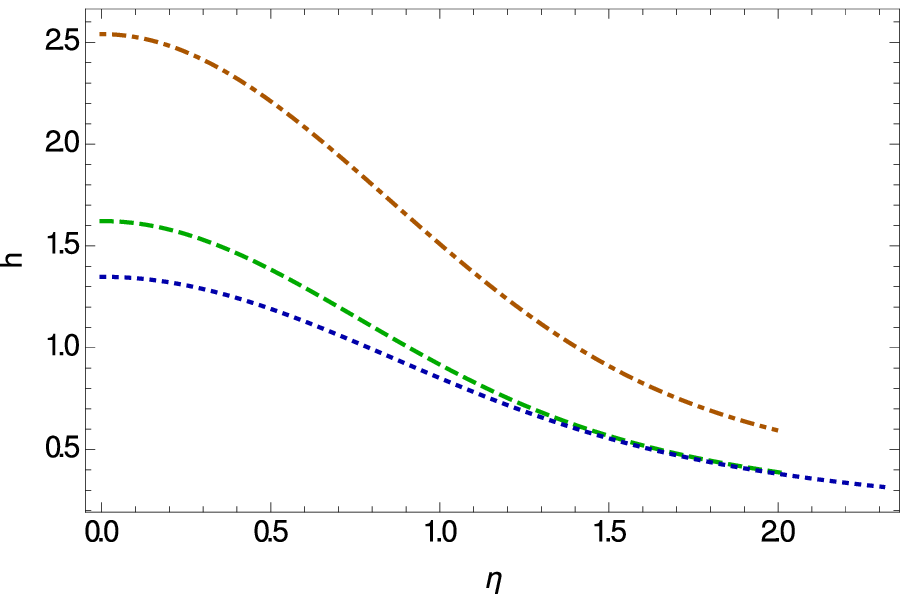}\hspace{.4cm}
	\includegraphics[width=8.0cm]{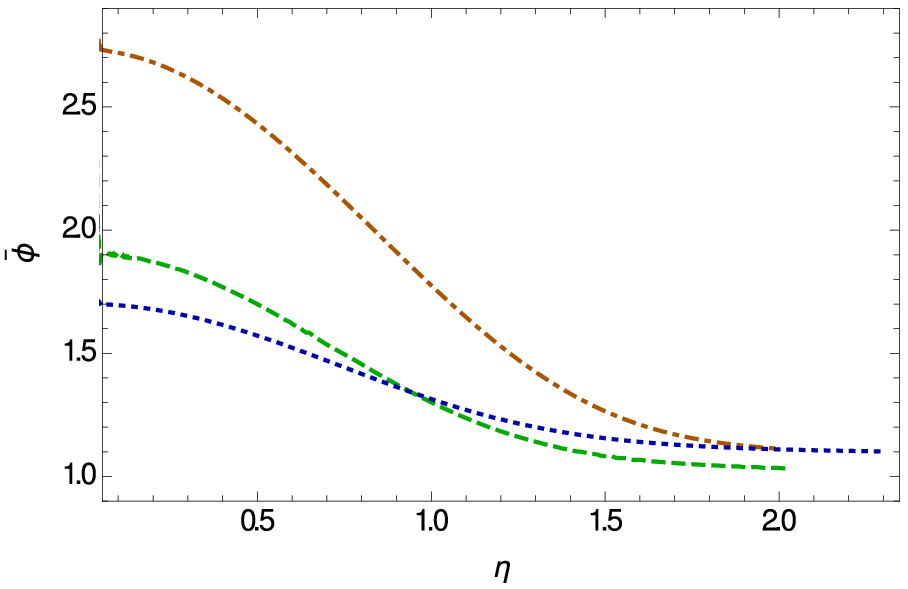}
	\caption{Variation of the scaled temporal component of Weyl vector field $h$ (left panel) and of the scalar field $\bar{\phi }$ (right panel) inside the stiff fluid star in Weyl geometric gravity as a function of the radial distance from the center of the  star $\eta$ for three different values of the
		constants$\bar{\alpha}$, $\bar{\xi}$ and $\bar{\gamma}$: $\bar{\alpha}=0.5$, $\bar{\xi}=0.05$  and  $\bar{\gamma}=0.76$  (dashed curve), $\bar{\alpha}=0.55$,  $\bar{\xi}=0.59$  and  $\bar{\gamma}=1.1$  (dotted curve), and $\bar{\alpha}=0.3$,  $\bar{\xi}=0.19$  and  $\bar{\gamma}=0.31$ (dot-dashed curve).   For all cases the central density is $4.9\times 10^{15}\; {\rm g/cm}^3$.}
	\label{stiff-vec}
\end{figure*}

The variations of the temporal component of the Weyl vector, and of the scalar field inside the stiff fluid star are depicted in Fig.~\ref{stiff-vec}. Both $h$ and $\bar{\phi}$ reach their maximum values at the center of the star, and they decrease rapidly towards the surface of the star. The scalar field generally tending towards a constant value near the vacuum boundary.

\begin{figure}[htbp]
	\centering
	\includegraphics[width=8.0cm]{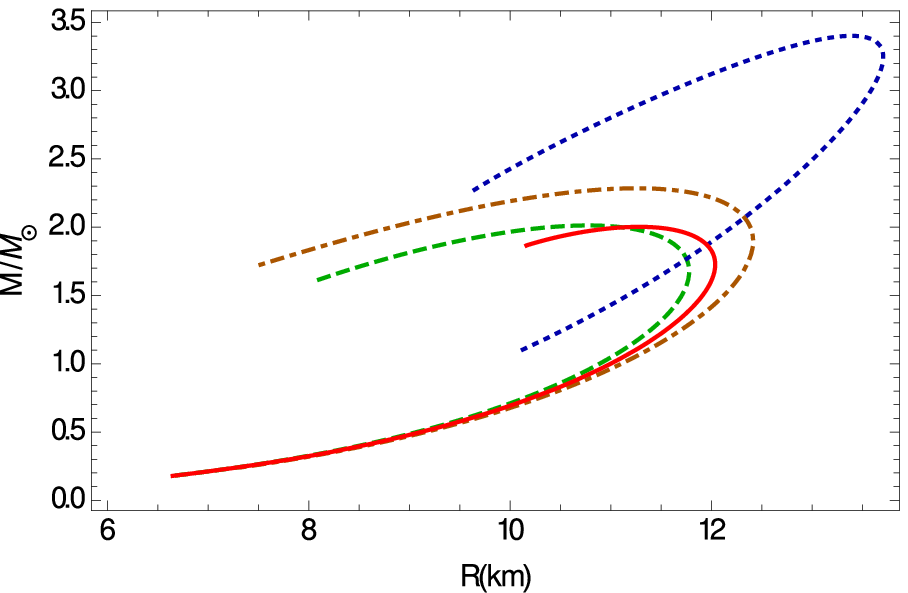}
	\caption{The mass-radius relation for stiff fluid stars in Weyl geometric gravity for three different values of the	constants $\bar{\alpha}$, $\bar{\xi}$ and $\bar{\gamma}$:  $\bar{\alpha}=0.5$, $\bar{\xi}=0.05$  and  $\bar{\gamma}=0.76$  (dashed curve), $\bar{\alpha}=0.55$,  $\bar{\xi}=0.59$  and  $\bar{\gamma}=1.1$  (dotted curve), and $\bar{\alpha}=0.3$,  $\bar{\xi}=0.19$  and  $\bar{\gamma}=0.31$ (dot-dashed curve). The solid curve represents the standard general relativistic  mass-radius relation for stiff stars. }
	\label{stiff-mr}
\end{figure}

The mass-radius relation for stiff fluid stars in Weyl geometric gravity is plotted, for various values of the model parameters,  in Fig.~\ref{stiff-mr}. Compact stellar configurations, with maximum masses of the order of $3.5M_{\odot}$ are possible even for central densities of the order of $\rho_c=5\times 10^{15}\;{\rm g/cm}^3$, when the corresponding maximum mass of the general relativistic star is only $2M_{\odot}$. Hence, Weyl geometric effects can lead to a significant increase in the mass of the equilibrium configurations of massive neutron stars.

Finally, in Table~\ref{stiff-tab} we present a number of selected mass and radii values for Weyl geometric stars, with central densities of the order of $10^{15}\;{\rm g/cm}^3$. The maximum masses essentially depend on the numerical values of the model parameters, leading, with an appropriate choice,  to mass values that can exceed the general relativistic limit of $3.2M_{\odot}$.

\begin{table}[h!]
	\begin{center}
		\begin{tabular}{|c|c|c|c|}
			\hline
			$\bar{\alpha}$&~~~$0.3$~~~&~~~$0.5$~~~&~~~$0.55$~~~ \\
			\hline
			$\bar{\xi}$ &~~~$0.19$~~~&$~~~0.05~~~~$&$~~~0.59~~~~$ \\
			\hline
			$\bar{\gamma}$ &~~~$0.31$~~~&$~~~0.76~~~~$&$~~~1.1~~~~$ \\
			\hline
			\quad$M_{max}/M_{\odot}$\quad& $~~~2.28~~~$& $~~~2.01~~~$& $~~~3.4~~~$\\
			\hline
			$~~~R\,({\rm km})~~~$& $~~~11.26~~~$& $~~~10.77~~~$& $~~~10.18~~~$\\
			\hline
			$~~~\rho_{c} \times 10^{-15}\,({\rm g/cm}^3)~~~$& $~~~2.10~~~$& $~~~2.46~~~$& $~~~1.04~~~$\\
			\hline
		\end{tabular}
		\caption{The maximum masses and the corresponding radii  and central densities for the stiff fluid stars in Weyl geometric gravity.}\label{stiff-tab}
	\end{center}
\end{table}

\subsection{Radiation fluid stars}

The radiation fluid equation of state plays a major role in astrophysics. Such an equation of state can describe the dense core of neutron stars, assumed to consist of cold degenerate (non-interacting) fermions \cite{Gle1, Gle2, Gle3}. Moreover, self-gravitating high density photon stars, obeying the radiation fluid equation of state, could also exist \cite{Ch1,Ch2,Ch3,Ch4}. Stars made of a radiation fluid could possible exist even in Newtonian gravity \cite{Ch5}.

In the following we will investigate the properties of radiation fluid stars in Weyl geometric gravity. For the equation of state of the matter inside the star we adopt the radiation fluid equation of state,
\begin{align}
p=\frac{1}{3}\rho.
\end{align}

This equation of state is the high density limit of the isothermal spheres, obeying the linear barotropic equation of state $p=\gamma \rho$,  with $\gamma ={\rm constant}$. It also follows from the limiting condition $T^{(m)}=0$, satisfied by the trace of the matter energy-momentum tensor once the condition of the positivity of trace is imposed.

From the consistency condition of the Weyl current, using Eq.~\eqref{tr3} with the radiation fluid equation of state,  we obtain for the temporal component of the Weyl vector and for the coupling constant $\gamma $ the following relation,
\begin{align}
 \gamma h^2=0.
\end{align}

Hence, in the following, we set $\bar{\gamma}=0$, so that the Weyl consistency condition is automatically satisfied. In this case, we use the vector field equation to obtain its evolution of the Weyl vector inside the star. The initial condition  used in the numerical calculations to solve the evolution equation of the Weyl vector  are $h(0)=h_c=0.89$ and $h'_c=-0.6\times10^{-3}$. For simplicity, we also set $\bar{\alpha}=0$, and consider the different values of $\bar{\xi}$.
The stop point in integration is $\rho=2.7\times 10^{14}\;{\rm g/cm}^3$. The range of central density in the numerical integration is between $3.0\times 10^{14}\, {\rm g/cm}^3$, and $1.1\times 10^{16}\, {\rm g/cm}^3$. For this set of initial values, the maximum mass of the standard general relativistic stars is $M=1.75M_\odot$, with radius $R=10.94\,{\rm km}$, corresponding to a central density $\rho_c=2.52\times 10^{15}\,{\rm g/cm}^3$.

\begin{figure*}[htbp]
	\centering
	\includegraphics[width=8.0cm]{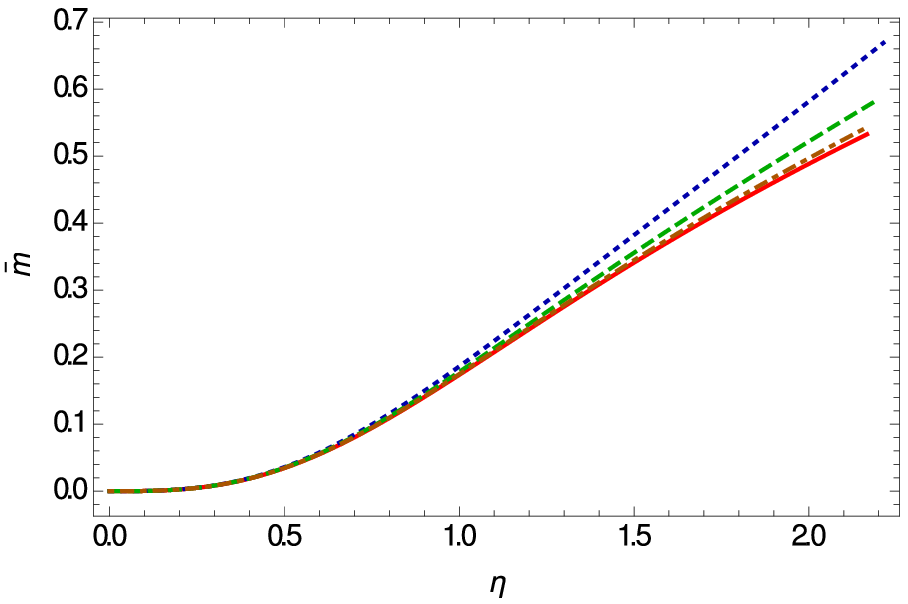}\hspace{.4cm}
	\includegraphics[width=8.0cm]{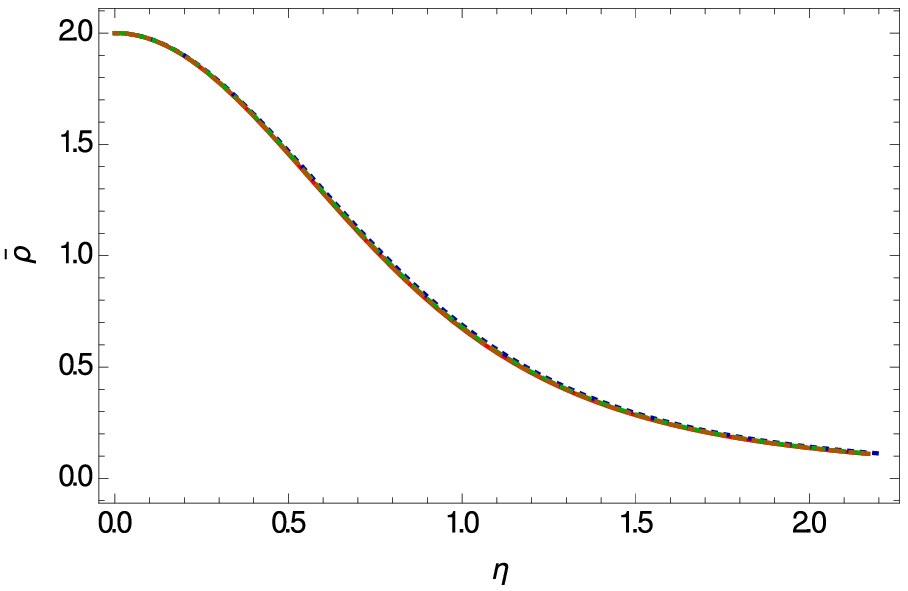}
	\caption{Variation of the interior mass (left panel)  and density (right panel) profiles of the radiation fluid star in Weyl geometric gravity as a function of the radial distance from the center of the  star $\eta$ for three different values of the constant $\bar{\xi}$:  $\bar{\xi}=0.29$ (dashed curve), $\bar{\xi}=0.49$  (dotted curve), and   $\bar{\xi}=0.15$ (dot-dashed curve). The solid curve represents the standard general relativistic  mass and density profile for radiation fluid stars. For all cases, the central density is $\rho _c=4.9 \times 10^{15}\; {\rm g/cm}^3$. }
	\label{rad-mass-dens}
\end{figure*}

The variations inside the star of the mass and density profiles is shown in Fig.~\ref{rad-mass-dens}. The Weyl geometric effects do not have a significant influence  on the density profile, as compared to the general relativistic case. The density monotonically decreases towards the vacuum boundary of the star, thus allowing the definition of the radius of the star. However, Weyl geometric effects can be seen on the mass of the radiation fluid star, leading to a significant increase of the total mass.

\begin{figure*}[htbp]
	\centering
	\includegraphics[width=8.0cm]{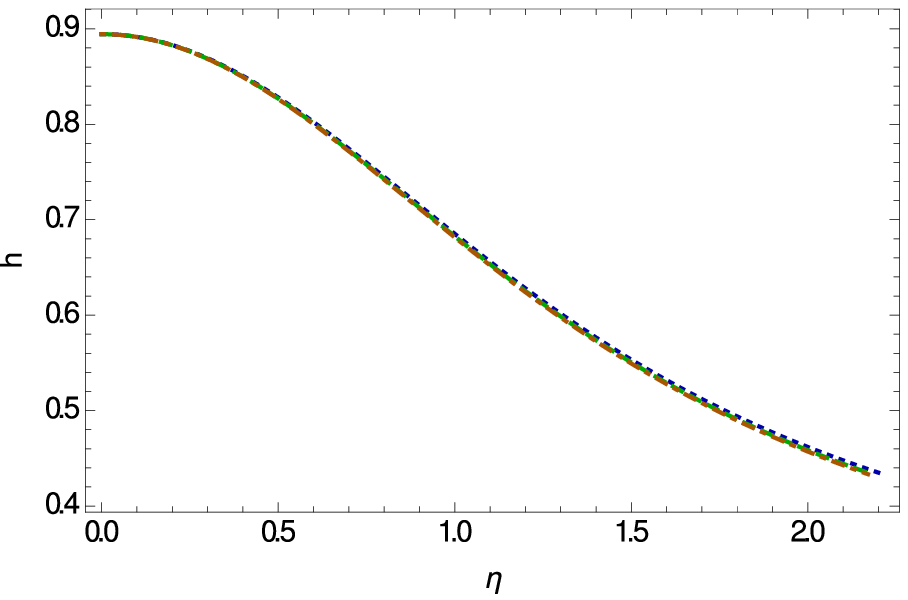}
	\includegraphics[width=8.0cm]{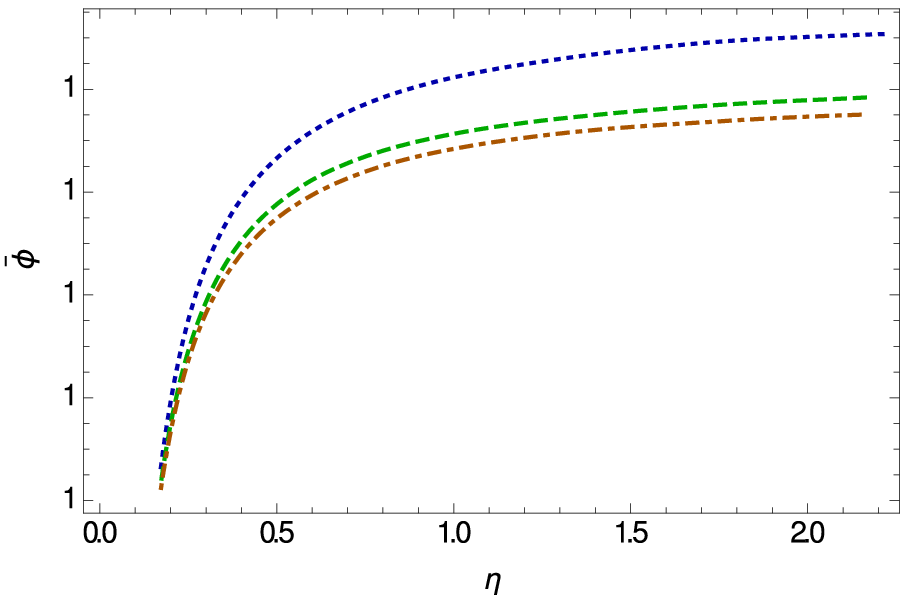}
	\caption{Variation of the scaled temporal component of Weyl vector field $h$ (left panel) and of the scalar field $\bar{\phi }$ (right panel) inside the radiation fluid star in Weyl geometric gravity as a function of the radial distance from the center of the  star $\eta$ for three different values of the
		constant $\bar{\xi}$:  $\bar{\xi}=0.29$ (dashed curve), $\bar{\xi}=0.49$  (dotted curve), and   $\bar{\xi}=0.15$ (dot-dashed curve).   For all cases the central density is $\rho_c=4.9\times 10^{15}\; {\rm g/cm}^3$.}
	\label{rad-vec}
\end{figure*}

The variations of the Weyl vector and of the scalar field are presented in Fig.~\ref{rad-vec}. The Weyl vector reaches its maximum value at the center of the star, and it decreases rapidly towards the star's surface. Its evolution is basically independent on the numerical values of the parameter $\bar{\xi}$. On the other hand, the scalar field is an increasing function of the radial coordinate, reaching its maximum value near the surface of the star. The variation of the scalar field is strongly influenced by the numerical values of $\bar{\xi}$.

\begin{figure}[htbp]
	\centering
	\includegraphics[width=8.0cm]{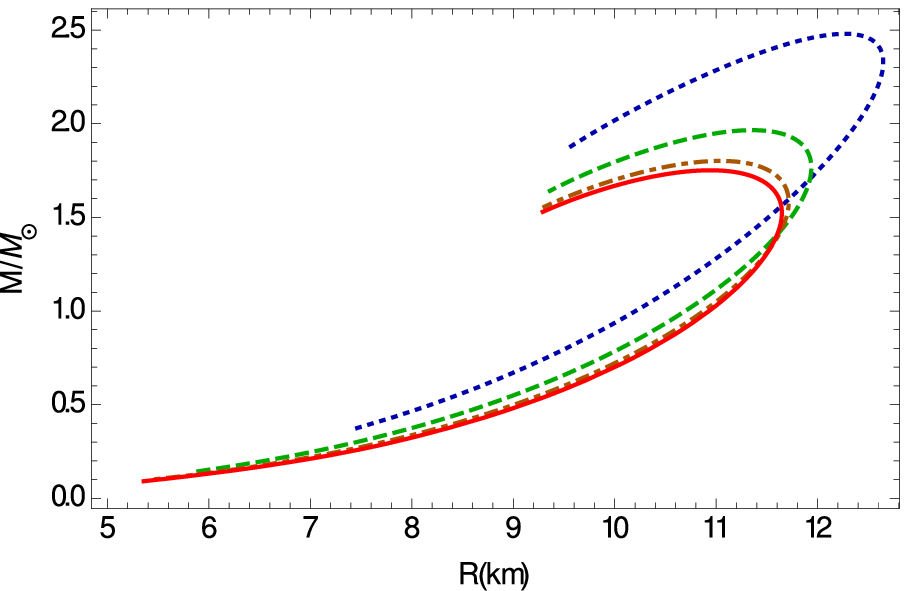}
	\caption{The mass-radius relation for radiation fluid stars in Weyl geometric gravity,  for three different values of the constant  $\bar{\xi}$: $\bar{\xi}=0.29$ (dashed curve), $\bar{\xi}=0.49$  (dotted curve), and   $\bar{\xi}=0.15$ (dot-dashed curve). The solid curve represents the standard general relativistic  mass-radius relation for radiation fluid stars. }
	\label{rad-mr}
\end{figure}

The mass-radius relation of the radiation fluid stars in Weyl geometric gravity is presented in Fig.~\ref{rad-mr}. Equilibrium structures having maximum masses higher than in standard general relativity can be achieved in Weyl geometric gravity for stars obeying the radiation fluid equation of state. The increase in mass is significant, if for general relativistic stars the maximum mass is around $1.7M_{\odot}$, the maximum mass of a radiation fluid star can reach in Weyl gravity values of the order
of $2.5M_{\odot}$.

A selected set of mass and radius values of radiation fluid stars in Weyl geometric gravity are presented in Table~\ref{rad-tab}.

\begin{table}[h!]
	\begin{center}
		\begin{tabular}{|c|c|c|c|}
			\hline
			$\bar{\xi}$ &~~~$0.15$~~~&$~~~0.29~~~~$&$~~~0.49~~~~$ \\
			\hline
			\quad$M_{max}/M_{\odot}$\quad& $~~~1.80~~~$& $~~~1.97~~~$& $~~~2.48~~~$\\
			\hline
			$~~~R\,({\rm km})~~~$& $~~~11.04~~~$& $~~~11.35~~~$& $~~~12.28~~~$\\
			\hline
			$~~~\rho_{c} \times 10^{-15}\,({\rm g/cm}^3)~~~$& $~~~2.37~~~$& $~~~1.98~~~$& $~~~1.19~~~$\\
			\hline
		\end{tabular}
		\caption{The maximum masses and the corresponding radii and central densities for the radiation fluid stars in Weyl geometric gravity.}\label{rad-tab}
	\end{center}
\end{table}

\subsection{Quark stars with MIT bag model equation of state }

A hadron-quark phase transition, taking place in the dense cores of the neutron star, is considered as a realistic possibility for the formation of quark matter \cite{Gle3}. Moreover, as shown by many theoretical studies, strange quark matter, consisting of the $u$ (up), $d$ (down) and $s$ (strange) quarks is the most energetically favorable state of baryon matter \cite{Witten}. The possibility of the existence of stars made of quarks was initially proposed in \cite{It} and \cite{Bod}, respectively, and further developed from an astrophysical perspective in  \cite{Al2} and \cite{Al2c}, respectively. There are two possible ways for the formation of strange matter. The first is the cosmological quark-hadron phase transition taking place in the early Universe, while the second is the conversion of neutron matter into strange matter at ultrahigh densities in neutron stars, thus leading to the formation of quark stars \cite{Gle3}.

By assuming that the interactions of quarks and gluons are small, the energy density $\rho $ and the pressure $p$ of a quark-gluon plasma can be calculated by using finite temperature quantum field theoretical methods. By neglecting quark masses,  the equation of state of the quark - gluon plasma is \cite{Wein}
\begin{eqnarray}
\rho &=&\left( 1-\frac{15}{4\pi }\alpha _{s}\right)
\frac{8\pi ^{2}}{15}T^{4}+N_{f}\left( 1-\frac{50}{21\pi
}\alpha_{s}\right) \frac{7\pi ^{2}}{10} T^{4}
    \nonumber\\
&&+ \sum_{f}3\left( 1-2\frac{\alpha _{s}}{\pi }\right) \left( \pi
^{2}T^{2}+\frac{\mu
_{f}^{2}}{2}\right) \frac{\mu _{f}^{2}}{\pi ^{2}}+B,  \label{s4}
\end{eqnarray}
or, equivalently,
\begin{equation}
\rho =\sum_{i=u,d,s}\rho _{i}+B,
\label{s5}
\end{equation}%
where $\alpha _s$ is the strong interaction coupling constant, $T$ is the temperature,  $\mu _{f}$ is the chemical potential, while $B$, the bag constant, is the difference between the energy density of the perturbative and non-perturbative QCD vacuum. The thermodynamic parameters of the quark-gluon plasma are related by the equation of state of the quark matter,  given by
\begin{equation}\label{s7}
p+B=\sum_{i=u,d,s}p_{i},
\end{equation}
 or
\begin{equation}\label{s6}
3p= \rho -4B.
\end{equation}

The entropy density $s$ of the quark-gluon plasma is given by
$s=\left( \partial p/\partial T\right) _{\mu }$. From a physical point of view, Eq.~(\ref{s6}) gives the equation of state of a system of massless
particles, in the presence of corrections originating in the QCD trace anomaly, and
the perturbative interactions. These corrections  are always negative. For example, for $\alpha _{s}=0.5$, the energy density of the quark gluon plasma
 at a given temperature is about two times smaller than the energy density of the gas of massless particles \cite{Wein}.

In the following, we will investigate the astrophysical properties of strange quark stars, described by the MIT bag model equation of state (\ref{s6}) in the framework of Weyl geometric gravity. In this case, by using the equation of state (\ref{s6}) in the consistency condition \eqref{tr3}, it turns out that the function $h^2=-2B/\gamma$ is constant. Moreover,  to have positive values of $h^2$, the coupling parameter $\gamma$ must be negative. In the following we set $B=1.03\times 10^{15}\; {\rm g/cm}^3$. In standard general relativity, the maximum mass of a quark star is $2 M_{\odot}$, with a radius of $10.92\,{\rm km}$, corresponding to a central density $\rho_c=1.98 \times 10^{15}\;{\rm g/cm}^3$.

\begin{figure*}[htbp]
	\centering
	\includegraphics[width=8.0cm]{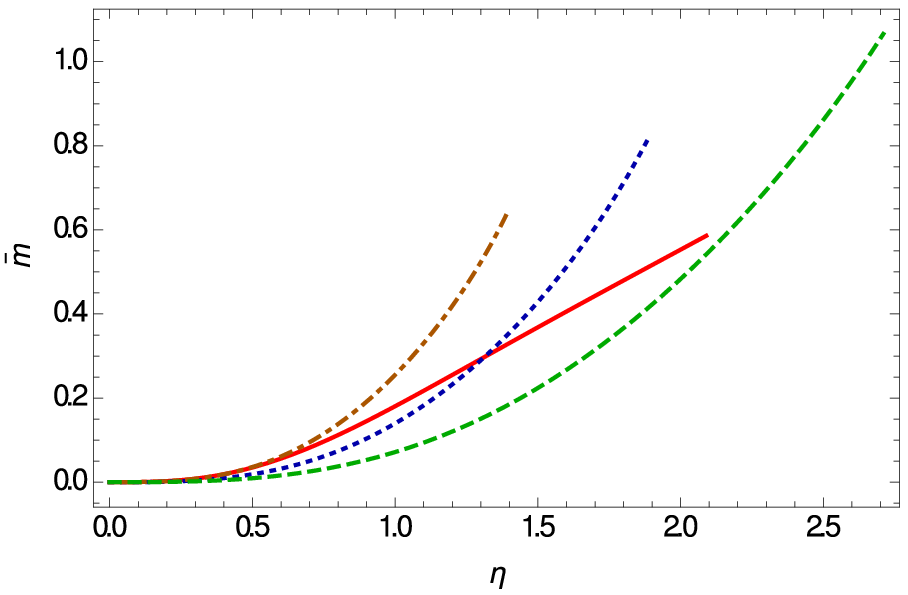}\hspace{.4cm}
	\includegraphics[width=8.0cm]{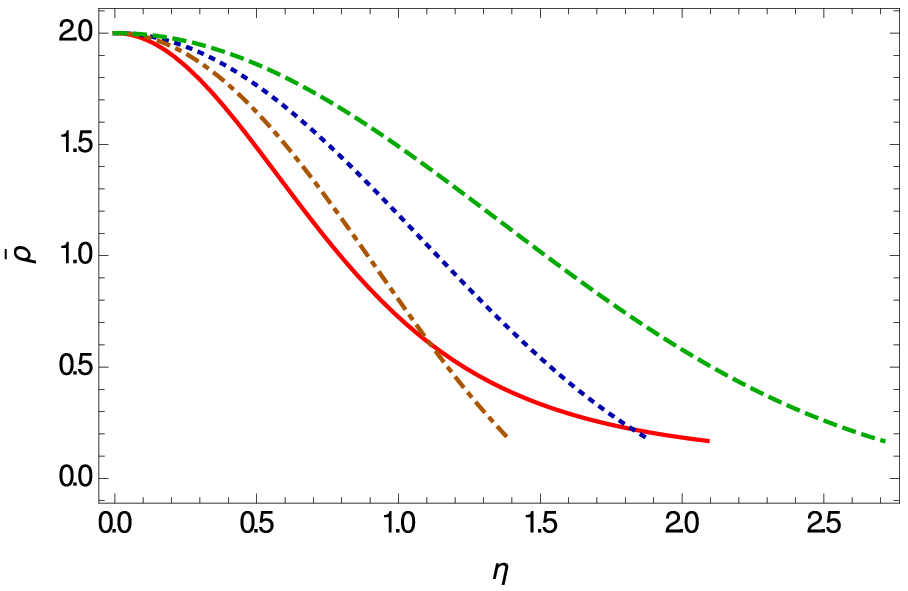}
	\caption{Variation of the interior mass (left panel) and density (right panel) profiles of MIT quark star in Weyl geometric gravity as a function of the radial distance from the center of the  star $\eta$ for three different values of the
		constants $\bar{\alpha}$, $\bar{\xi}$ and $\bar{\gamma}$:  $\bar{\alpha}=0.1$, $\bar{\xi}=0.15$  and  $\bar{\gamma}=-0.06$  (dashed curve), $\bar{\alpha}=0.15$,  $\bar{\xi}=0.39$  and  $\bar{\gamma}=-0.2$  (dotted curve), and $\bar{\alpha}=0.2$,  $\bar{\xi}=0.98$  and  $\bar{\gamma}=-0.4$ (dot-dashed curve). The solid curve represents the standard general relativistic  mass and density profile for MIT quark stars. For all cases the central density is $4.9\times 10^{15}\; {\rm g/cm}^3$. }
	\label{MIT-mass-dens}
\end{figure*}

The variations of the interior mass and density profiles of the quark stars are plotted in Fig.~\ref{MIT-mass-dens}. The pressure exactly vanishes on the vacuum boundary of the star, thus leading to a unique radius of the quark star. The mass and the density distributions strongly depend on the parameters of the Weyl geometric gravity model, leading to the presence of a large variety of internal structures of the stars.

\begin{figure}[htbp]
	\centering
	\includegraphics[width=8.0cm]{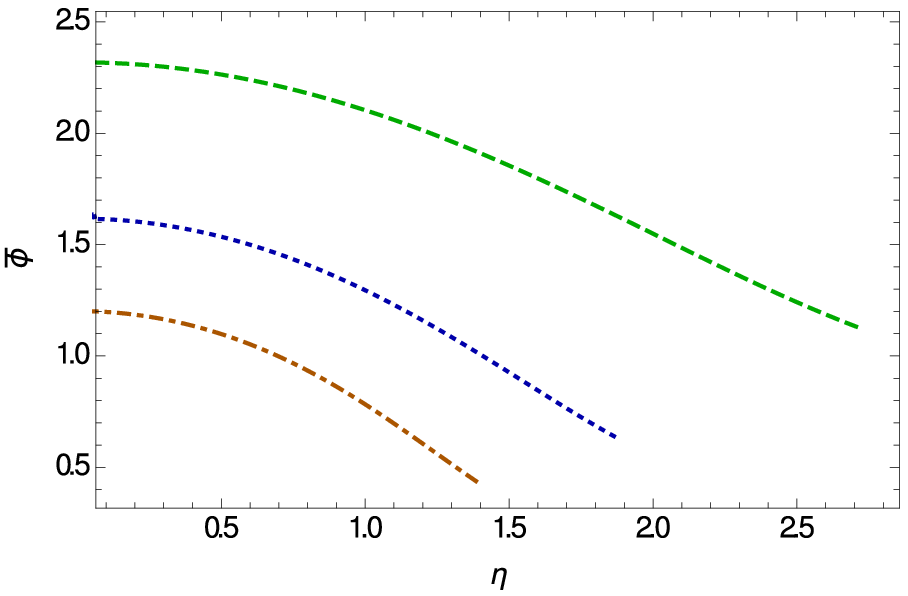}
	\caption{of Variation of the scalar field $\bar{\phi }$ in Weyl geometric gravity inside the MIT quark star as a function of the radial distance from the center of the  star $\eta$ for three different values of the
		constants $\bar{\alpha}$, $\bar{\xi}$ and $\bar{\gamma}$:    $\bar{\alpha}=0.1$, $\bar{\xi}=0.15$  and  $\bar{\gamma}=-0.06$  (dashed curve), $\bar{\alpha}=0.15$,  $\bar{\xi}=0.39$  and  $\bar{\gamma}=-0.2$  (dotted curve), and $\bar{\alpha}=0.2$,  $\bar{\xi}=0.98$  and  $\bar{\gamma}=-0.4$  (dot-dashed curve).  For all cases the central density is $4.9\times 10^{15}\; {\rm g/cm}^3$.}
	\label{MIT-vec}
\end{figure}

The behavior of the scalar field inside the star is depicted in Fig.~\ref{MIT-vec}. The scalar field is a monotonically decreasing function of $\eta$, having its maximum value at the center of the star. The variation of the field is significantly influenced by the adopted range of model parameters $\bar{\alpha}$, $\bar{\xi}$ and $\bar{\gamma}$, respectively.

\begin{figure}[htbp]
	\centering
	\includegraphics[width=8.0cm]{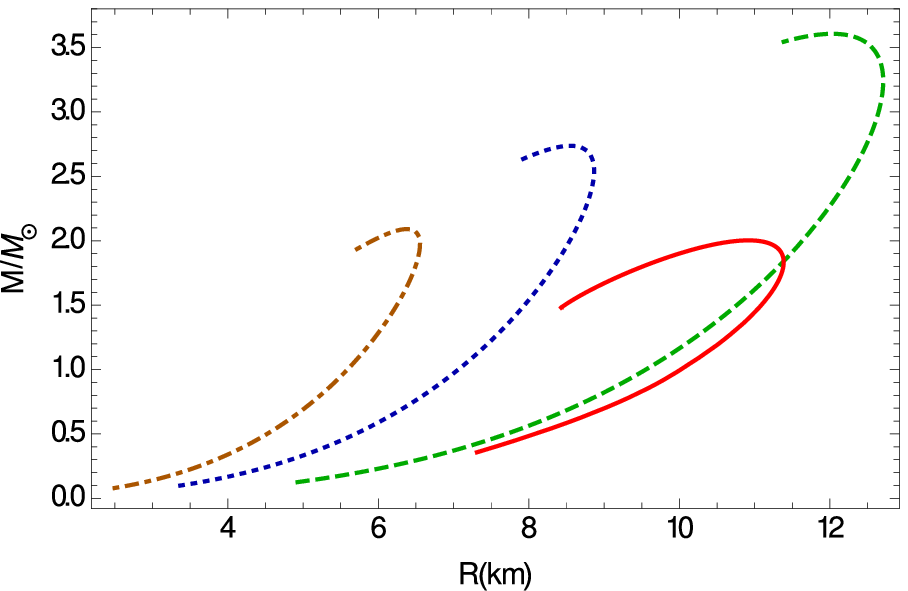}
	\caption{The mass-radius relation for MIT quark stars in Weyl geometric gravity for three different values of the	constants $\bar{\alpha}$, $\bar{\xi}$ and $\bar{\gamma}$:   $\bar{\alpha}=0.1$, $\bar{\xi}=0.15$  and  $\bar{\gamma}=-0.06$  (dashed curve), $\bar{\alpha}=0.15$,  $\bar{\xi}=0.39$  and  $\bar{\gamma}=-0.2$  (dotted curve), and $\bar{\alpha}=0.2$,  $\bar{\xi}=0.98$  and  $\bar{\gamma}=-0.4$  (dot-dashed curve). The solid curve represents the standard general relativistic  mass-radius relation for MIT quark stars. }
	\label{MIT-mr}
\end{figure}

The mass-radius relation for quark stars in the Weyl geometric gravity theory is represented in Fig.~\ref{MIT-mr}. As one can see from the Figure, the presence of the Weyl vector and of the scalar field lead to significant increases of the maximum of the equilibrium configurations of quark stars.

 A selected sample of maximum masses of the quark stars in the Weyl geometric gravity theory is shown, for different values of the model parameters, in Table \ref{MIT-tab}.

\begin{table}[h!]
	\begin{center}
		\begin{tabular}{|c|c|c|c|}
			\hline
			$\bar{\alpha}$&~~~$0.15$~~~&~~~$0.10$~~~&~~~$0.20$~~~ \\
			\hline
			$\bar{\xi}$ &~~~$0.39$~~~&$~~~0.15~~~~$&$~~~0.98~~~~$ \\
			\hline
			$\bar{\gamma}$ &~~~$-0.20$~~~&$~~~-0.06~~~~$&$~~~-0.40~~~~$ \\
			\hline
			\quad$M_{max}/M_{\odot}$\quad& $~~~2.73~~~$& $~~~3.61~~~$& $~~~2.09~~~$\\
			\hline
			$~~~R\,({\rm km})~~~$& $~~~8.55~~~$& $~~~12.03~~~$& $~~~6.38~~~$\\
			\hline
			$~~~\rho_{c} \times 10^{-15}\,({\rm g/cm}^3)~~~$& $~~~10.2~~~$& $~~~12.6~~~$& $~~~7.69~~~$\\
			\hline
		\end{tabular}
		\caption{The maximum masses, and the corresponding radii and central densities for the MIT bag model quark stars in Weyl geometric gravity.}\label{MIT-tab}
	\end{center}
\end{table}

\subsection{Bose-Einstein Condensate stars}

When in a bosonic system the temperature drops below a certain critical value, a phase transition does occur, with the particles occupying the same quantum ground state. The corresponding system of particles is called  a Bose-Einstein Condensate (BEC). From a physical point of view,  a BEC corresponds to the presence of a sharp peak in the phase space \cite{Chav}. The critical temperature of the BEC transition is given by $T_{cr}=\left(2\pi \hbar ^2/mk_B\right)n^{2/3}$, where $m$ is the mass of the particle in the condensate, $\hbar$ is Planck's constant, $k_B$ is Boltzmann's constant, and $n$ is the number density of the particles. The possibility of the existence of Bose-Einstein Condensation processes has also been investigated in nuclear physics. For example, at very high densities matter exists in a form of a degenerate Fermi gas of quarks. If the attractive interaction is enough strong,  at some critical temperature the fermions may condense into the bosonic zero mode, forming a Bose-Einstein quark condensate \cite{Chav}. Hence, the possibility of the existence of some forms of Bose-Einstein Condensates in high density neutron stars has a strong support from the results of nuclear physics  (see \cite{Gle3} for a detailed discussion of this problem). The non-relativistic theory of the Bose-Einstein Condensate stars is based on the hydrodynamic representation of the Gross-Pitaevskii equation \cite{Chav}. The radius of the BEC star, in which the neutron have formed Cooper pairs,  is given by $R=6.61\times \left(a/1\;{\rm fm}\right)^{1/2}\times \left(m/2m_n\right)^{-3/2}$ km, where $a$ is the scattering length, and $m_n$ is the neutron mass \cite{Chav}. For $a=1$ fm, the radius of the star is around 7 km. The mass of the star is obtained as $M=1.84\times \left(\rho_c/10^{16}\;{\rm g/cm^3}\right)\times \left(a/1\;{\rm fm}\right)^{3/2}\times \left(m/2m_n\right)^{-9/2}M_{\odot}$. For $a=1$ fm, the mass of the star is of the order of $M=0.92M_{\odot}$. The equation of state of a Bose-Einstein Condensate is given by \cite{Chav},
\begin{align}
p=k\rho^2,
\end{align}
where the constant $k$ is given by $k=0.1856\times 10^5\times \left(a/1\;{\rm fm}\right)\times \left(m/2m_n\right)^{-3}\;{\rm cm^5/g\;s^2}$. For a discussion of the general relativistic effects on the astrophysical parameters of the BEC stars see \cite{Chav}. For a Bose-Einstein Condensate the Weyl consistency condition, given by Eq.~\eqref{tr3} can be written as
\begin{align}
h^2=\frac{1}{2\gamma}\left(3k\rho-1\right)\rho,
\end{align}

For positive values of $\gamma$, the expression in the parenthesis in the right hand side of the above equation should be always greater or equal to zero. The initial value of $h$ is obtained as $h^2(0)=(1/2\gamma)\left(3k\rho_c-1\right)$.  Hence, we set the stop point in the numerical integration at $\rho=1/3k$. In following we consider $\bar{k}=\rho_c k=0.4$, and thus the stop point in the numerical integration is $2.04\times 10^{15}\, g/cm^3$. The range of central density in the numerical solution is considered in the range  $2.2\times 10^{15}\, {\rm g/cm}^3$ and $6.74\times 10^{15}\, {\rm g/cm}^3$, respectively. In this case the maximum mass of the standard general relativistic BEC stars is $M=2M_\odot$, with radius $R=11.17\,{\rm km}$, with a central density $\rho_c=2.58\times 10^{15}\,{\rm g/cm}^3$.

\begin{figure*}[htbp]
	\centering
	\includegraphics[width=8.0cm]{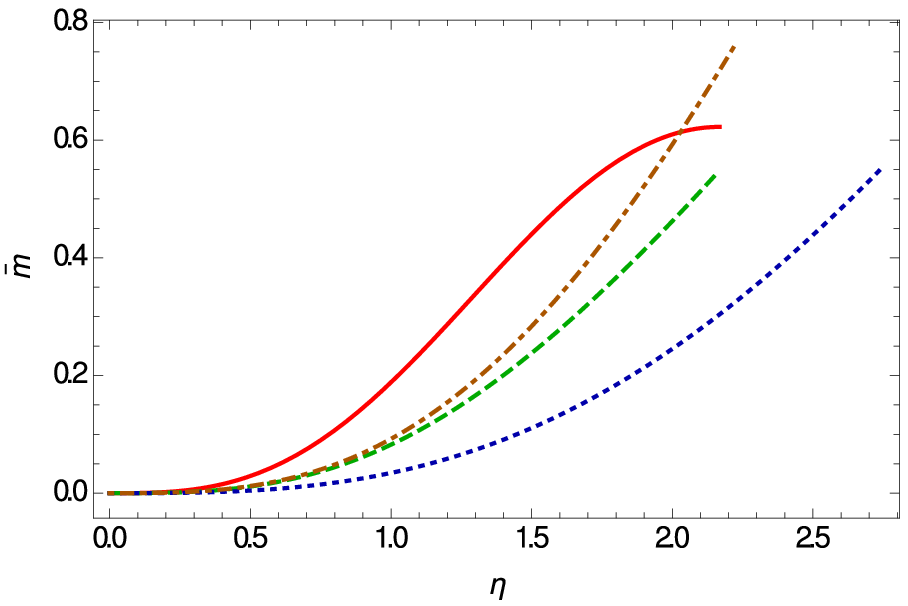}
	\includegraphics[width=8.0cm]{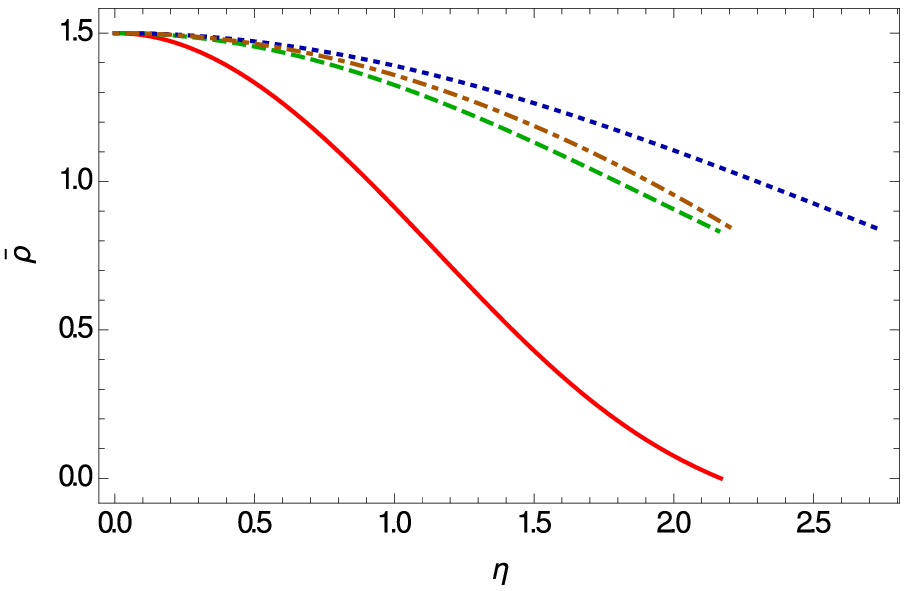}
	\caption{Variation of the interior mass (left panel) and density (right panel) profiles of Bose-Einstein Condensate stars in Weyl geometric gravity  as a function of the radial distance from the center of the  star $\eta$ for three different values of the constants $\bar{\alpha}$, $\bar{\xi}$ and $\bar{\gamma}$: $\bar{\alpha}=0.074$, $\bar{\xi}=0.19$  and  $\bar{\gamma}=0.03$  (dashed curve), $\bar{\alpha}=0.07$,  $\bar{\xi}=0.05$  and  $\bar{\gamma}=0.06$  (dotted curve), and $\bar{\alpha}=0.12$,  $\bar{\xi}=0.34$  and  $\bar{\gamma}=0.10$ (dot-dashed curve).  The solid curve represents the standard general relativistic  mass and density profile for Bose-Einstein condensate stars. For all cases the central density is $\rho_c=3.67\times 10^{15}\; {\rm g/cm}^3$. }
	\label{BEC-mass-dens}
\end{figure*}

The mass and density profiles of the BEC stars in Weyl geometric gravity are presented, for a selected sample of model parameters, in Fig.~\ref{BEC-mass-dens}. The mass profile depends strongly on the model parameters, whose variation leads to a large number of possible stellar structures. More important differences do appear in the behavior of the matter density. While in standard general relativity one can uniquely define a zero density vacuum boundary, the Weyl consistency condition leads to BEC stars having a non-zero surface density, and a non-zero pressure.

\begin{figure*}[htbp]
	\centering
	\includegraphics[width=8.0cm]{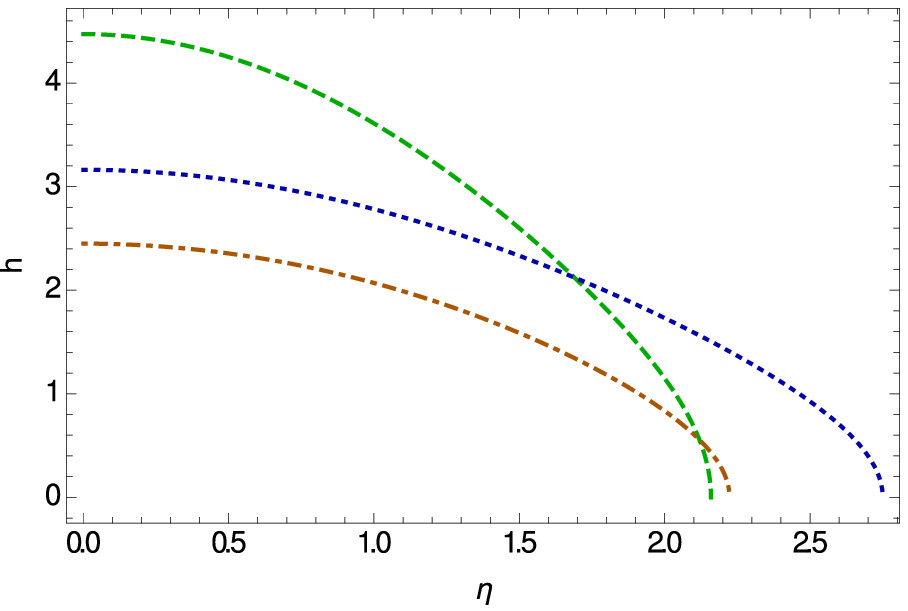}
	\includegraphics[width=8.0cm]{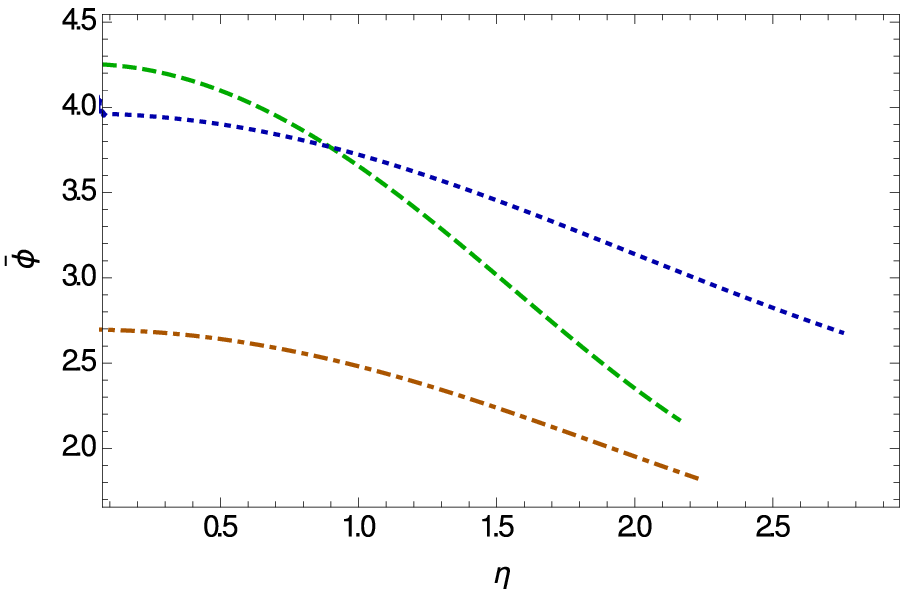}
	\caption{Variation of the scaled temporal component of Weyl vector field $h$ (left panel) and of the scalar field $\bar{\phi }$ (right panel) in Weyl geometric gravity inside the Bose-Einstein Condensate stars as a function of the radial distance from the center of the  star $\eta$ for three different values of the
		constants$\bar{\alpha}$, $\bar{\xi}$ and $\bar{\gamma}$:  $\bar{\alpha}=0.074$, $\bar{\xi}=0.19$  and  $\bar{\gamma}=0.03$  (dashed curve), $\bar{\alpha}=0.07$,  $\bar{\xi}=0.05$  and  $\bar{\gamma}=0.06$  (dotted curve), and $\bar{\alpha}=0.12$,  $\bar{\xi}=0.34$  and  $\bar{\gamma}=0.10$ (dot-dashed curve).   For all cases the central density is $\rho_c=3.67\times 10^{15}\; {\rm g/cm}^3$.}
	\label{BEC-vec}
\end{figure*}

The variations of the temporal component of the Weyl vector and of the scalar field inside the star are represented in Fig.~\ref{BEC-vec}. The Weyl vector reaches its maximum value at the center of the star, decreases towards its surface, and vanishes for a given value of $\eta$. The scalar field is a monotonically decreasing function of $\eta$, having finite values on the star's surface.

\begin{figure}[htbp]
	\centering
	\includegraphics[width=8.0cm]{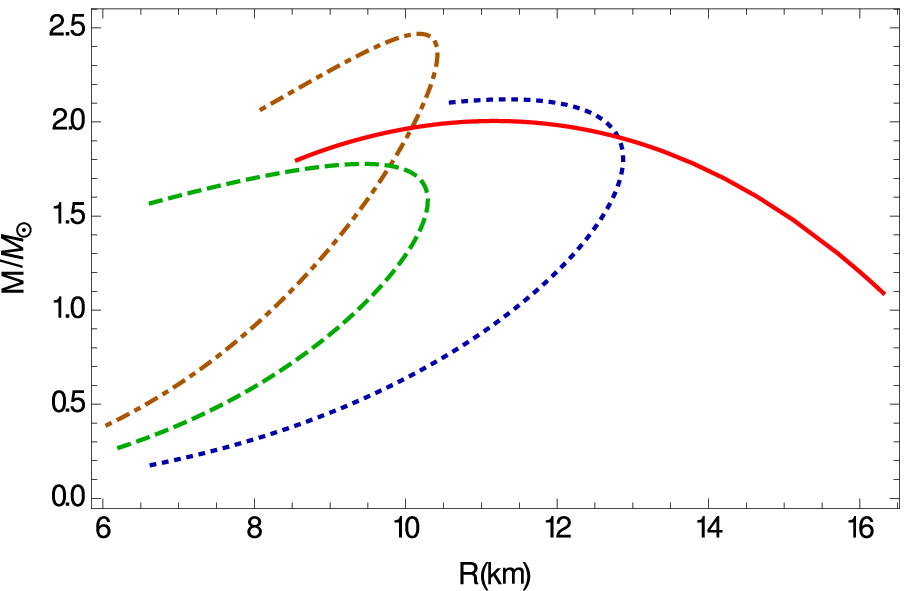}
	\caption{The mass-radius relation for Bose-Einstein Condensate stars in Weyl geometric gravity for three different values of the constants $\bar{\alpha}$, $\bar{\xi}$ and $\bar{\gamma}$:  $\bar{\alpha}=0.074$, $\bar{\xi}=0.19$  and  $\bar{\gamma}=0.03$  (dashed curve), $\bar{\alpha}=0.07$,  $\bar{\xi}=0.05$  and  $\bar{\gamma}=0.06$  (dotted curve), and $\bar{\alpha}=0.12$,  $\bar{\xi}=0.34$  and  $\bar{\gamma}=0.10$ (dot-dashed curve). The solid curve represents the standard general relativistic  mass-radius relation for Bose-Einstein condensate stars. }
	\label{BEC-mr}
\end{figure}

The mass-radius relations for BEC stars in Weyl geometric gravity are represented, for different values of the model parameters, in Fig.~\ref{BEC-mr}. Weyl geometric effects do allow for the existence of higher stellar masses for BEC stars.

Finally, some specific astrophysical parameters of BEC stars in Weyl geometric gravity are presented, for different values of the model parameters,  in Table~\ref{BEC-tab}.

\begin{table}[h!]
	\begin{center}
		\begin{tabular}{|c|c|c|c|}
			\hline
			$\bar{\alpha}$&~~~$0.07$~~~&~~~$0.074$~~~&~~~$0.12$~~~ \\
			\hline
			$\bar{\xi}$ &~~~$0.05$~~~&$~~~0.19~~~~$&$~~~0.34~~~~$ \\
			\hline
			$\bar{\gamma}$ &~~~$0.06$~~~&$~~~0.03~~~~$&$~~~0.10~~~~$ \\
			\hline
			\quad$M_{max}/M_{\odot}$\quad& $~~~2.12~~~$& $~~~1.70~~~$& $~~~2.47~~~$\\
			\hline
			$~~~R\,({\rm km})~~~$& $~~~11.36~~~$& $~~~9.44~~~$& $~~~10.18~~~$\\
			\hline
			$~~~\rho_{c} \times 10^{-15}\,({\rm g/cm}^3)~~~$& $~~~5.12~~~$& $~~~4.35~~~$& $~~~4.17~~~$\\
			\hline
		\end{tabular}
		\caption{The maximum masses and the corresponding radii and central densities for the Bose-Einstein Condensate stars in Weyl geometric gravity.}\label{BEC-tab}
	\end{center}
\end{table}

\section{Discussions and final remarks}\label{sect4}

 The study and the analysis of astrophysical properties of compact objects are of fundamental importance for our understanding of gravitational theories, since these stellar type structures  provide a very good opportunity for the investigation of the properties of dense matter in extreme conditions, and for the study of the  strong gravity regime. Moreover, a self-consistent modified or extended theory of gravity must also have a significant effect on the properties of stellar type objects. Thus, a modified theory of gravity must have important effects not only on galactic and cosmological scales, but also on the properties of stable compact stellar objects, such as black holes, neutron stars, or white dwarfs.

 A large number of accurate observations on the  masses and radii obtained from massive pulsars, the gravitational wave event GW170817, or PSR J0030+0451
mass-radius relation from  Neutron Star Interior Composition Explorer (NICER) data have dramatically changed our understanding of the mass distribution of neutron stars, by challenging the paradigm according to which the mass of the neutron stars peaks at around the Chandrasekhar limit of $1.4M_{\odot}$ \cite{Horv}. For example, the mass of the pulsar PSR J0952-0607, the fastest known rotating neutron star in the disk of the Milky Way, has been determined recently, giving a maximal mass of the order of $2.52 M_{\odot}$ ($M = 2.35 \pm  0.17M_{\odot}$). The mass value of PSR J0952-0607 definitely represents a challenge for our understanding of the equation of state of the dense-matter. One possible explanation for this, and other similarly high, mass values may be related to the existence of some forms of exotic matter in the interior of these stars. However, an equally reasonable explanation for the mass distribution of massive neutron stars could be related to the presence of modified gravity effects, which must show up in the strong gravity limits of the theory.

In the present paper we have investigated the implications of the conformally invariant theories on the stellar structure. We have adopted the simplest theoretical model, initially proposed by Weyl \cite{Weyl, Weyl1}, which has been reformulated as a scalar-vector-tensor theory in \cite{Gh4,Gh5,Gh6,Gh7}. This reformulation has deep physical implications, also allowing to establish a close relation between gravity and elementary particle physics, via the Stueckelberg mechanism. We have also added an effective  matter term to the gravitational action, assumed to be a function of the standard matter Lagrangian of the ordinary matter, and of the square of the Weyl vector. In order to implement the conformal invariance of the gravitational field equations, the trace condition has also been imposed, which relates the trace of the effective energy-momentum tensor with the covariant divergence of the Weyl current. For the expression of the Lagrangian we have adopted the simplest possible form, by assuming that $\mathcal{L}$ is the sum of the ordinary matter Lagrangian $L_m$, and the square of the Weyl vector, $\omega _{\mu}\omega ^\mu =\omega ^2$. Of course, other functional forms of the effective Lagrangian are possible, and some of these forms may be obtained by using some input from the elementary particle physics.

After obtaining the static spherically symmetric field equations, we have considered several classes of stellar models, corresponding to specific choices of the equation of state of the dense matter. We have investigated in detail, by numerically integrating the field equations, constant density stars, stiff fluid, radiation fluid, quark and BEC stars, respectively, by fully taking into account in each case the trace condition. We have thus constructed classes of astrophysical objects, showing significant differences with respect to their general relativistic counterparts. The main difference is related to the masses of the Weyl geometric gravity objects, which are significantly higher as compared to the corresponding general relativistic objects. For the stiff fluid, radiation fluid, quark and BEC equations of state masses of the order of $2.5-2.6M_{\odot}$ can be easily achieved, for certain specific combinations of model parameters. Hence, Weyl geometric gravity can described (at least) the masses of the pulsars  B1957+20 (the Black Widow pulsar) ($M=1.6-2.4M_{\odot}$)\cite{Black},  PSR J2215+5135 ($M = 2.27_{-0.15}^{+0.17} M_{\odot}$) \cite{Lin}, PSR J1614-2230 ($M=2.01\pm 0.04M_{\odot}$) \cite{Ant}, PSR J0952-0607 ($M=2.35\pm 0.17M_{\odot}$) \cite{Rom}, and of the secondary object in the GW190814 event ($M=2.50-2.67M_{\odot}$) \cite{Ligo}, respectively. The effects of the Weyl geometric gravity theory on the global  properties are that the stars become more massive, and with larger radii, thus leading to larger surface redshifts, and compactness.

Of course, the properties of stellar objects in Weyl geometric gravity essentially depend on the equation of state of dense matter. Testing modified theories of gravity requires a prior knowledge of the equation of state of the dense matter, and thus, constraints independent of the gravitational theories could play a crucial role in the understanding of the properties of the neutron stars. In \cite{Lope1} it was proposed to use the function  we propose the function $\left.L\right|_{\varepsilon}\left(\epsilon _H)\right)=P_H\left[\left(\varepsilon _E-\varepsilon_H\right)/\left(\varepsilon _E\varepsilon _H\right)\right]$, the indices $E$ and $H$ refer to the exotic and hadronic phases, respectively,   as a diagnostic tool for constraining the equation of state of dense matter in a first order phase transition. The existence of a family of latent-heat maxima
to constrain the equation of state of neutron matter, is relevant even in modified theories of gravity \cite{Lope1}. Moreover, latent heat is a significant characteristic of the equation of state in cold QCD.  When deriving the equation of state of matter one uses the standard general relativistic formalism for the interpretation of the astrophysical data, and thus general relativity is essentially included in the equation of state. Hence, such the use of such equations of state may be problematic in modified gravity \cite{Lope2}. For the astrophysical applications of modified gravity it is important to provide equations of state coming from nuclear and particle physics, which are independent of general relativity. Such a family of equations of state was provided in \cite{Lope2},  and they can be used to constrain modified gravity theories. These equations of State, rely only on first principle approaches, including causality, thermodynamic stability, and perturbation theory.

The effects of modified gravity on the equation of state of dense matter have been considered extensively in the physical literature. In \cite{Kul}, the thermodynamic properties of an ideal quantum gas in a modified gravity theory with Lagrangean $L_g=\left(R-2\Lambda\right)/16\pi G+\alpha R^2+\beta R_{\mu \nu}R^{\mu \nu}+\gamma R_{\mu \nu \tau \sigma}R^{\mu \nu \tau \sigma}$, with $\alpha, \beta, \gamma$ constants, were considered. The dependence of the Fermi energy and of the chemical potential on the curvature was also obtained.  The surface of a star in Eddington-inspired Born-Infeld modified gravity was studied in \cite{Kim}, by assuming a polytropic equation of state. In order to avoid the presence of singularities at the surface of the star,  the gravitational backreaction on the particles is considered, which leads to an extended polytropic equation of state of the form $p=K\rho ^{\gamma}+\epsilon \rho^{3/2}$, with $K$, $\Gamma$, $\epsilon$ constants. Due to the increase in pressure, the surface of the star is no longer singular. The role of non-metricity in quantum fields was investigated in \cite{Del} by considering the 4-fermion contact interactions. The scale of non-metricity was constrained to be greater than 1 TeV, a result that follows from the analysis of the Bhabha scattering. The analysis was carried out by considering modified theories of gravity in the metric-affine approach.

It was pointed out in \cite{Hoss1}  that the Tolman-Oppenheimer-Volkoff equations for neutron stars are usually solved by
equations of states that are obtained in the Minkowski spacetime. On the other hand, the equations of state that are obtained in a curved spacetime
also include the effects of the gravitational time dilation, which is a consequence of the radial variation of the interior metric in the star.  This effect leads to much higher
mass limits for neutron stars. For example, in the $\sigma-\omega$ model of nuclear matter, the maximum mass increases from $1.61M_{\odot}$ to $2.24M_{\odot}$, respectively. The effects of the time dilation on the equation of state inside slowly rotating neutron stars was investigated in \cite{Hoss2}, where it was shown that the equation of state also depends on the frame dragging effect. However, while the time dilation effect leads to a significant increase of the mass of the star, the frame dragging has a negligible influence on the maximum mass of the star. On the other hand, the angular momentum of the star increases in the presence of an equation of state that takes into account the curvature of the space-time.

 The equations of state of a Fermi gas were derived, by maximizing the Fermi-Dirac entropy, by considering the Palatini $f(R)$ gravity,  in the relativistic and nonrelativistic limits in \cite{Woj}. As a main result of this investigation it was found that to obtain a consistent description of the microphysical properties one must use the specific theory of gravity under consideration.

When numerically integrating the gravitational field equations in static spherical symmetry in $f(R)$ type modified theories of gravity,  models containing the $R^2$ term may lead to some unphysical features of the solutions. For example, in the $f(R)=R+\alpha R^2$ model, which allows the presence of heavier stars than in general relativity, it was found that there are regions where the enclosed mass decreases with the radius \cite{Or}. Similar results were obtained in \cite{Ash} for the same model, where it was shown that the stellar mass bounded by the surface of the star decreases when the value of $\alpha$ increases, and that the scalar curvature does not tend to zero at the surface of the star, but it increases exponentially. The problem of the matching in the $f(R)$ gravity theories was considered in \cite{Resco}, where it was shown that the calculation of the mass of the neutron star requires a careful matching  between interior and exterior solutions, firstly at the star's edge, and secondly at large radii, where the Newtonian potential is used to identify the mass of the neutron star. A possible solution to this problem may be scalarization \cite{S1,S2,S3}, a process similar to a phase transition, occurring after some physical parameters characterizing a compact object, like, for example, the compactness or spin, exceed a critical value. Hence, as a results of scalarization, massive compact objects are endowed with a scalar field configuration. For a detailed discussion of the scalarization see the recent review \cite{S4}.  Scalarization may also provide an explanation for the fact new fundamental fields have not been detected by the present observations. However, future high precision astrophysical observations may allow their detection.

In order to obtain a full consistency check of the interior solutions obtained in the present work it is necessary to match them with exterior solutions of the field equations. The vacuum field equations of the Weyl geometric gravity have been obtained numerically in \cite{C4} for several configurations of the Weyl vector. A physically realistic solution would require a smooth matching between the interior and exterior solutions. The requirement for the exterior solution of passing the Solar System tests would then impose strong constraints not only on the vacuum solution, but also on the parameters of the interior stellar model.

In Weyl geometric gravity,  the geometric effects and the contributions of the effective scalar and vector fields dominate, and their contribution to the total matter-energy balance is the main cause of the significant increase of the mass.  Many stellar mass black hole candidates, have also been observed, having masses between $3.8$ and 6 Solar masses. Seven of them have masses bigger than $5M_{\odot}$. On the other hand, the size of the stellar mass black hole population in our galaxy is estimated to be of the order of 100 millions \cite{Kov}. Since in Weyl geometric gravity theory stellar masses of the order of 5-6$M_{\odot}$ are also possible, some, if not all, stellar black hole candidates may be in fact Weyl geometric stars.

One possibility to test the possible existence of Weyl geometric stars is through the study  of the astrophysical properties of the thin accretion disks existing around many neutron stars and black holes. The radiation properties of the accretion disks around Weyl geometric stars, and black holes, and general relativistic black holes and stars may be different, since the emission from the disk takes place in the strong gravity regime.  Hence, the emissivity properties of the stars, and of their accretion disk, may provide the crucial signature to differentiate Weyl geometric gravity  stars from general relativistic stars, and  black holes.

Weyl geometric gravity compact objects posses a very complicated internal structure, which is associated to an equally complicated stellar dynamics. We will consider in a future publication several astrophysical signatures of Weyl geometric gravity stars, which would allow to differentiate between this type of stars, and their counterparts predicted by general relativity, or other modified gravity theories.

\section*{Acknowledgments}

 The work of TH is supported by a grant of the Romanian Ministry of Education and Research, CNCS-UEFISCDI, project number PN-III-P4-ID-PCE-2020-2255 (PNCDI III).

\end{document}